 \documentclass[11pt,twocolumn]{article}

 \pdfoutput=1

  \usepackage[final]{pdfpages} 
\usepackage[round]{natbib}   

\usepackage[english]{babel}

\usepackage{graphicx}

\usepackage{amsmath}
\usepackage{chemformula} 

\usepackage{enumitem}

\usepackage{color}

\usepackage{float}
\usepackage{ifthen}

\usepackage{hyperref}

\usepackage{titlesec}

\titlespacing*{\section}{0pt}{1.1\baselineskip}{\baselineskip}

\begin{document}

\def\hmath#1{\text{\scalebox{1.5}{$#1$}}}
\def\lmath#1{\text{\scalebox{1.4}{$#1$}}}
\def\mmath#1{\text{\scalebox{1.2}{$#1$}}}
\def\smath#1{\text{\scalebox{.8}{$#1$}}}

\def\hfrac#1#2{\hmath{\frac{#1}{#2}}}
\def\lfrac#1#2{\lmath{\frac{#1}{#2}}}
\def\mfrac#1#2{\mmath{\frac{#1}{#2}}}
\def\sfrac#1#2{\smath{\frac{#1}{#2}}}

\def\pow{^\mmath}



\twocolumn[

\begin{center}
{\bf \Large {The absolute seawater entropy: Part~I. Definition}}
\\
\vspace*{3mm}
{\Large by Dr.Hab. Pascal Marquet}. \\
\vspace*{3mm}
{\large Retired from M\'et\'eo-France (CNRM), Toulouse, France.}
\\ \vspace*{2mm}
{\large  \it E-mail: pascalmarquet@yahoo.com}
\\ \vspace*{2mm}
{\large URL: \url{https://sites.google.com/view/pascal-marquet}}
\\ \vspace*{2mm}
{\Large\bf  \color{black} Accepted in: }
{\Large\bf  \color{black} \it\bf 
{Comptes Rendus Geoscience (Paris, France).}}
\vspace*{2mm}\\
{\large  \color{black} {Initial submission: 
                       7th of October 2024}.}
\vspace*{1mm}\\
{\large  \color{black} Three Reviewers' comments received 
                       the 10th of September, 2025.}
\vspace*{1mm}\\
{\large  \color{black} First revised version: 
                       30th of November, 2025.} 
\vspace*{1mm}\\
{\large  \color{black} Second Reviewers' comments received 
                       the 11th of February, 2026.}
\vspace*{1mm}\\
{\large  \color{black} {\bf Accepted version:
                       23th of March, 2026} \;\;\;
                 \url{https://doi.org/10.5802/crgeos.333}.} 
\vspace*{1mm}
\end{center}

\begin{center}
{\large \bf Abstract}
\end{center}
\vspace*{-3mm}

\hspace*{7mm}
The absolute entropy of seawater is defined as an improved version of the relationship defined by 
Millero in 1976 and 1983.
The first improvements concern the complex non-linear dependence of entropy on pressure, temperature and salinity, with the use of the standard TEOS10 formulation based on a fit of the oceanic Gibbs function to more recent observations.
On the other hand, more recent thermodynamic tables have been used to increase the accuracy of the Millero's salinity increment to this standard formulation, to deduce the absolute version of entropy with new values for the pure-water and sea-salts absolute reference entropies.
The differences between the values of the seawater entropy calculated with the Millero and TEOS10 formulations (standard and absolute) are documented, before a more complete study shown in the second part of the paper of the absolute seawater entropy computed from observed vertical profiles and analysed surface datasets.
\vspace*{0mm}

\begin{center}

First-Review Answers to the Editors and Reviewers 
can be found on Zenodo \citep{Marquet_Zenodo_2025_Answers_R1}.
\vspace*{2mm}

Supplementary materials are provided in the Zenodo file
\citet{Marquet_Zenodo_2025_Sup_Mat_3rd_law}. 
\vspace*{2mm}

Second-Review Answers to the Reviewers 
can be found on Zenodo \citep{Marquet_Zenodo_2026_Answers_R2}.
\vspace*{2mm}
\end{center}

] 

 \section{Introduction}
\label{section_introduction}
\vspace*{-2mm}

The absolute entropy of the moist-air atmosphere was first defined 
by \citet{Hauf_Hoeller_1987} but was never calculated 
or studied for its own sake until 
recently$\,$\footnote{$\:$In the studies of 
\citet{Marquet2011QJ,Marquet14,MarquetGeleyn2015,Marquet2017JAS,
MarquetThibaut2018JAS,MarquetBechtold2020,Marquet_al_2022} 
and \citet{Marquet_Stevens_JAS22}.}.
Similarly, the absolute entropy of sea-salt oceans has already 
been defined in the papers by 
\citet[][hereafter ML76]{Millero_Leung_1976} and 
\citet[][hereafter M83]{Millero_1983}, 
recalled in \citet{Sharqawy_Lienhard_Zubair_2010} and 
\citet{Qasem_Thermo_Prop_Saline_Water_2023} but not in \citet{Nayar_al_2016}, 
without any attempt until now to calculate it from in-situ measurements 
or from numerical model outputs.

These ``absolute'' definitions of the entropies of the atmosphere and the ocean are obtained from the reference values of entropies available in all thermodynamic and thermochemical 
tables$\,$\footnote{$\:$In particular: \citet[][]{Kelley_1932,Latimer_al_1938,
Rossini_al_1952,
Laidler_1956,Lewis_Randall_1961,
Robinson_Stokes_Electrolyte_1970,Robie_al_1978,Wagman_al_NBS_1982,
Grenthe_al_NEA_TDB_1992,
Gokcen_Reddy_1996,
Chase_1998,Atkins_Paula_2014,
Grenthe_al_NEA_TDB_2020} and \citet[][]{Atkins_Paula_Keeler_2023}.},
this for the various species 
\ch{N2}, \ch{O2}, \ch{Ar}, \ch{CO2} and \ch{H2O}  
making up the moist-air atmosphere, 
and for liquid water, {Na${}^{+}$}, {Mg${}^{2+}$}, ...,   
{Cl${}^{-}$}, {SO${}_4^{2-}$}, ...
for the seasalt ocean.

These absolute reference values have been computed by applying 
the third law of thermodynamics, which, following the works of 
\citet{Nernst_1906} and then 
\citet{Planck1911,Planck1917}, 
stipulates that the entropies of the most stable crystalline 
state of all bodies cancel out at the absolute zero of temperature.
This third law is used for the calorimetric calculations
$S(T)=S(0)+\int_0^T c_p(T')\: d\ln(T')+\sum_k L_k(T_k)/T_k$
based on $S(T=0~\mbox{K})=0$, with the integral of the specific heat capacities 
$c_p(T)$ and including the impact of all phase changes at temperatures 
$T_k$ with the latent heats $L_k(T_k)$.
The third law also corresponds to the statistical calculations 
$S(T)=0+k\: \ln(W)$, where 
the $0$ term indicates that no other arbitrary term must be added, 
where $k$ is the Planck-Boltzmann constant, and where $W$, 
the number of quantum configurations, 
must be evaluated at the absolute temperature $T$ for all translational (Sackur-Tetrode-Planck), 
rotational, and vibrational degrees of freedom of atoms and molecules. 
Note that if the original 
third law corresponds to both $S(T=0~\mbox{K})=0$ in the calorimetric formula 
and to the $0$ term in the statistical one, this original 
definition of \citet{Planck1911,Planck1917} 
must be nowadays amended by considering possible residual entropies $S(0) \neq 0$ at $0$~K in 
the calorimetric formula only (but not the statistical one) 
and for a few species like \ch{H2O} (ice Ih), due to the 
next computations of \citet{Pauling1935} and \citet{Nagle1966}.

Differently, almost all other definitions of the quantities called ``entropies'' 
in atmospheric and oceanic studies deviate from these methods  
by choosing other reference values that differ from 
both the calorimetric and statistical-quantum values 
provided by the third law of thermodynamics.  
More precisely, the alternative definitions are obtained by 
arbitrarily cancelling out these reference entropy values 
at the zero Celsius temperature or the triple point at $0.01{}^{\circ}$C  
(instead of zero Kelvin) for dry air, liquid water, and 
sea salts$\,$\footnote{$\:$For instance, in the contributions by   
\citet{Iribarne_Godson_1973,IribarneGodson1981}, \citet{Emanuel1994}, 
\citet{Romps_2008}, \citet{Pauluis_Czaja_Korty_2010}, 
\citet{Raymond2013} and \citet{Mrowiec_al_2016}, 
among so many others, for the atmosphere, 
and in the contributions by 
\citet{Fofonoff_1962}, 
\citet{Fofonoff_JGR_1985}, \citet{Feistel_1993}, 
\citet{Feistel_Hagen_1995}, \citet{Wagner_Pruss_2002}, 
\citet{IAPWS_Feistel_2003}, \citet{Warren_Prog_Ocean_2006}, 
\citet{Feistel_al_2008}, \citet{Sun_al_Deep_Sea_Res_2008}  
\citet[][]{Feistel_TEOS_manual_2010}, ...,  
up to 
\citet{Feistel_2019},  
\citet{McDougall_al_2023} and 
\citet{Feistel_2024_OS20}, for the ocean.}.
The fact that the values of the reference entropies have no impact 
in most oceanographic applications, unless one wants to calculate 
and study the entropy of seawater itself, is a key reason for the 
dominant approach where these reference entropies are defined 
arbitrarily, as in TEOS10.

It is within this framework that the 
TEOS10\footnote{$\:$TEOS10 
in short for ``\,Thermodynamic Equation Of Seawater - 2010.\,''
See: \url{http://www.teos-10.org/publications.htm}} 
software \citep[][]{Feistel_TEOS_manual_2010} has been designed, 
with thermodynamic functions computed with greater consistency from 
the Gibbs function fitted on more numerous and more recent observed 
datasets than before.
However, although \citet[][p.~19]{Millero_History_seawater_2010} explained 
that ``{\it\, (...) new TEOS-10 (...) will be very useful to modelers 
examining the entropy and enthalpy of seawater,\,}''
this TEOS10's values can be amended to correspond to
the third law of thermodynamics and Millero's previous papers 
concerning the calculation of seawater entropy, due to arbitrary 
redefinitions made in TEOS10 of the reference entropies of liquid 
water and ocean salts.
Moreover, there have been a few papers, other than those already cited, 
where the absolute definitions of entropies have been explicitly mentioned 
or even taken into account: 
\citet[][]{Lemmon_al_2000} for the dry air;
\citet[][]{Feistel_Wagner_2005,Feistel_Wagner_2006} for the liquid water.

Consequently, the aim of the present paper is to compute and study the absolute seawater 
entropy state variable on concrete cases (observed vertical profiles and surface analyzed 
datasets, as shown in the next Part~II).
The chosen methodology described in Part~I consists of reconciling the TEOS10 formulation, 
with all its advantages, with the use of absolute values of reference entropies, 
as prescribed by Millero and available in all thermodynamic tables.

The paper is organized as follows.
I explain in section~\ref{section_seawater_entropy} 
the way the standard seawater entropy is presently 
computed in the reference TEOS10 software
(see subsection~\ref{subsection_standard_version}).
I then explained in the subsection \ref{subsection_need_update} 
why there is a need to update the computations of this standard seawater entropy, 
by taking into account the absolute value for the (pure) liquid-water entropy
recalled in the subsection~\ref{subsection_abs_entropies_water},
and also the absolute for the sea-salts entropies, first computed  
at $25{}^{\circ}$C in the subsection~\ref{subsection_abs_entropies_seasalts_25C}, 
and then at $0{}^{\circ}$C in the 
subsection~\ref{subsection_abs_entropies_seasalts_0C},
before to show in the subsection~\ref{subsection_abs_entropy_TEOS10}
how it is possible to modify the TEOS10 relationships  
to compute the seawater entropy for all conditions of 
temperature, salinity and pressure.

Several comparisons with observations and analysed surface datasets are conducted 
in the second part of the paper \citep{Marquet_CRAS_Geos_25_II}.
In this first theoretical part are shown in Section~\ref{section_numerical_applications} 
two numerical applications.
The impact of the salinity is studied in subsection~\ref{subsection_abs_entropy_Millero} 
with a comparison between the old Millero and other papers, including the standard TEOS10 
and the new absolute seawater entropy formulations. 
I show that several problems unfortunately prevent the use of the `relative' entropy 
formulation considered by Millero in 1976 and 1983. 
Moreover, this `relative' entropy formulation did not really take into account the 
absolute values of entropies and has never been used since then. 
This therefore justifies basing the present approach on the more modern formulation 
of TEOS10, with the sole addition of taking into account the absolute values 
of the reference entropies for liquid water and ocean salts.
The classic $t-S_{\rm A}$ oceanic diagrams are plotted in Subsection~\ref{subsection_t_S_diagram} 
by adding the new absolute isentropic (iso-entropy) lines to the isopycnic (iso-density) lines.   
Furthermore, I describe in subsection~\ref{subsection_General_impacts} 
other more general situations in physics where the 
absolute values of entropies impact different phenomena,  
with detailed computations shown in the Supplementary Materials  
\citep[][hereafter referred to as ``SM'']{Marquet_Zenodo_2025_Sup_Mat_3rd_law}, 
as well as answers to FAQs regarding the interest, or not, of studying the 
absolute entropies of the atmosphere and oceans.

I finally recall in the concluding Section~\ref{section_conclusion} 
the main results of the paper.

 \section{Computations of seawater entropy}
\label{section_seawater_entropy}
\vspace*{-2mm}

 \subsection{The standard TEOS10 present version}
\label{subsection_standard_version}
\vspace*{-2mm}

The aim of this section is to recall how the seawater entropy 
is calculated in the current standard version of TEOS10, 
and to show where the reference entropy values for liquid 
water and sea salts appear.

The specific 
entropy$\,$\footnote{$\:$It is of common use in oceanography to write the entropy with 
the symbols $\eta$ or $\sigma$ (instead of the symbol $S$ generally used 
in thermodynamics since Clausius, Boltzmann, and Planck) in order to keep 
the letters $S$, $S_{\rm P}$ or $S_{\rm A}$ for the salinity.
Note that other symbols like $\phi$ are also used to denote the entropy, 
with, for instance, the $T$-$\phi$ atmospheric diagram still called ``Tephigram'' 
in the UK and Canada.} 
$\eta$ called ``\,entropy of seawater\,'' is computed in 
\citet[][Eq.~2.10.1, p.~20]{Feistel_TEOS_manual_2010}$\,$\footnote{$\:$I will refer in the 
present paper to the pages and numbers of figures, equations, and tables of the version 
of the TEOS10's Manual downloaded in September 2024 from  
\url{http://www.teos-10.org/pubs/TEOS-10_Manual.pdf}} 
as a derivative of the specific seawater Gibbs' function $g$:  
\begin{align}
 \eta\:(S_{\rm A}, t, p) & \: = \;
 -\:\left.\frac{\partial g}{\partial t}\right|_{S_{\rm A},p}
 \: = \;
 -\: \frac{1}{40} \left.\frac{\partial g}{\partial y}\right|_{S_{\rm A},p}
 \label{eq_TEOS10_eta_dg_dt} \: ,
\end{align}
where $S_{\rm A}$ is the absolute salinity,
$t=T-273.15$ the Celsius temperature (with $T$ the absolute temperature), 
$p=P-P_{\rm SO}$ the ``sea pressure'' 
(Eq.~A.2.1, p.~73)
equal to the absolute pressure $(P)$ 
minus the standard ocean surface sea pressure 
$P_{\rm SO}=0.101\,325$~MPa 
(given in the table D.4, p.~145, and thus with $p$ close 
to the depth, in meters, when expressed in dbar),
and $y = t / (40{}^{\circ}$C$\,)$ a dimensionless temperature variable
(where $t_{\rm u}=40{}^{\circ}$C is given in the table D.4, p.~145).

The specific free enthalpy $g=G/m$ (with $m=1$~kg) of 
a multi-component dilute aqueous electrolyte solution 
was previously derived according to 
\citet[][Eq.~4.9, p.~266]{Feistel_Hagen_1995} 
from the practical osmotic factor considered in 
\citet[][Eq.~23-4, p.~334; hereafter LR61]{Lewis_Randall_1961} 
and \citet[][Eq.~116, p.~40]{Falkenhagen_Ebeling_electrolytes_1971}, 
leading to a formulation that can be rewritten as 
\begin{align}
 g & \:=\;   (1-C) \times \mu^0_{\rm w}\,(T,P)
         \:+\: C   \times \mu^0_{\rm s}\,(T,P)
  \nonumber \\
  &\quad
  \;+\; \sum_a \; \frac{X_a}{M_{\rm s}} \; R \; T\; C \; 
    \ln\!\left[\:  X_a\:\frac{C}{1-C}\:
                  \frac{M_{\rm w}}{M_{\rm s}}
       \:\right]
  \nonumber \\
  &\quad
  \:-\;
  \frac{\left[\:
    \sum_a\:N_a\:Z_a^2\:e^2/D(T,P)
  \:\right]^{3/2}}
  {\sqrt{\:36\:\pi\:\nu(T,P)\;k\;T\:N_{\rm w}\:}}
  \; ,
  \label{eq_FH95_4_9}
\end{align}
where: 
$C=1.00488\:S$ is the sea-salts concentration (i.e., the absolute salinity 
$S_{\rm A}/1000$); $M_{\rm w} \approx 18.0152$~g~mol\,${}^{-1}$ and $N_{\rm w}$ 
the molar mass and number of molecules of liquid water; 
$\mu^0_{\rm w}\,(T,P)$ the specific 
chemical function at infinite dilution for liquid water, 
$\mu^0_a\,(T,P)$, $X_a$, $M_a$ and $N_a$ the specific chemical function at infinite dilution,  
molar fraction, molar mass and number of molecules for each of the $n$ sea salts ($a=1,2,...,n$);  
$M_{\rm s} = \sum_a X_a \: M_a \approx 31.4058$~g~mol\,${}^{-1}$ the mean molar mass of sea salts
with $\sum_a X_a = 1$, 
$\mu^0_{\rm s}\,(T,P) = \sum_a \: X_a\: \mu^0_a\,(T,P) $ 
the mean specific sea-salts chemical function (at infinite dilution); 
and with the other terms in the third line described in 
\citet[][p.~264-266]{Feistel_Hagen_1995}.

According to \citet[][Eq.~2.6.1, p.~15]{Feistel_TEOS_manual_2010} 
the TEOS10 Gibbs function (\ref{eq_FH95_4_9}) is split in two 
and written as the sum of the 'pure water' ($W$) and 'salinity' 
($S$) parts 
\begin{align}
 g\:(x,y,z)
 & \: = \;
 g^{\rm W}_{\rm Fei03}\:(y,z)
 \; + \;
 g^{\rm S}_{\rm Fei08}\:(x,y,z)
 \label{eq_TEOS10_g_x_y_z} \: , 
\end{align}
with, from (\ref{eq_TEOS10_eta_dg_dt}),  
the entropy function written as
\begin{align}
 \eta\:(x,y,z)
 & \: = \;
 \eta^{\rm W}_{\rm Fei03}\:(y,z)  
 \; + \;
 \eta^{\rm S}_{\rm Fei08}\:(x,y,z) 
 \label{eq_TEOS10_eta_x_y_z} \: , 
\end{align}
both in terms of $y$ and the other dimensionless 
salinity and pressure variables:
$x^2 = S_{\rm P}/40 = S_{\rm A}/(40.188\,617\,$g~kg${}^{-1})$, 
where $S_{\rm P}$ is the practical salinity, and  
$z = p / (100$~MPa$\,)$ (see the table D.4, p.~145, and 
the Appendices G and H, p.~155-156).

The two Gibbs functions 
$g^{\rm W}_{\rm Fei03}\:(y,z)$ 
and $g^{\rm S}_{\rm Fei08}\:(x,y,z)$ 
were previously computed by  
\citet[][Table~10-27, p.~93]{IAPWS_Feistel_2003} and
\citet[][Table~17, p.~1666]{Feistel_IAPWS_2008}, 
respectively, as series expansions of $x^i\:y^j\:z^k$:
\begin{align}
 g^{\rm W}_{\rm Fei03}\:(y,z)  & \: = \;
   \sum_{j=0}^{7} \sum_{k=0}^{6} 
   \; g_{0jk} \:y^j\:z^k
   \label{eq_TEOS10_Gw_y_z} \: 
\end{align}
and 
\begin{align}
 g^{\rm S}_{\rm Fei08}\:(x,y,z)  & \: = \;
   \sum_{j=0}^{6} \sum_{k=0}^{5} \; [\: 
   \; g_{1jk} \; x^2\:\ln(x) 
 \nonumber \\
 & \quad\:+\: 
   \sum_{i=2}^{6} 
   \; g_{ijk} \:x^i \:] \:y^j\:z^k
 \label{eq_TEOS10_Gs_y_z} \: , 
\end{align}
with the TEOS10 coefficients $g_{ijk}$ listed in 
\citet[][Appendices G and H, p.~155-156]{Feistel_TEOS_manual_2010}.

According to (\ref{eq_TEOS10_eta_dg_dt}) 
applied to (\ref{eq_TEOS10_g_x_y_z}), and thus to 
(\ref{eq_TEOS10_Gw_y_z}) and (\ref{eq_TEOS10_Gs_y_z}),
the entropy coefficients corresponding 
to (\ref{eq_TEOS10_eta_x_y_z}) are 
$\eta_{\,ijk}=-\,(j+1)\:g_{i,\,j+1,\,k}/40$
and correspond to the relationships shown in   
Tables~\ref{Table_TEOS10_sigma_W_x_y_z}
and \ref{Table_TEOS10_sigma_S_x_y_z}, 
which enable us to compute directly the TEOS10 
seawater entropy without using the TEOS10 
software$\,$\footnote{$\:$Namely, as done in the 
part of the subroutine {\tt{gsw\_gibbs.f90}}
of the software {\tt{GSW-Fortran-3.05-6}}
with the derivation options
``\,{\tt{(ns.eq.0).and.(nt.eq.1).and.(np.eq.0)}}\,''.}.

The way in which the two parts of $g$ are defined 
in (\ref{eq_TEOS10_g_x_y_z}) can be understood by 
rewriting the first line of (\ref{eq_FH95_4_9})
using the definition 
$\mu^0_{\rm w} = h_{\rm w}-T\:\eta_{\rm w}$ and 
$\mu^0_{\rm s} = h_{\rm s}-T\:\eta_{\rm s}$ 
for the pure-liquid water and sea-salts chemical potentials, 
with the concentration $C={S_{\rm A}}/{1000}$ in kg~kg${}^{-1}$ 
depending on the absolute salinity $S_{\rm A}$ in g~kg${}^{-1}$.
Therefore, the first-order constant terms 
$\eta_{\rm w0}$ and $\eta_{\rm s0}$ (independent of $T$ and $P$) 
of the entropy $\eta$ impact the Gibbs function via the terms 
$(1-{S_{\rm A}}/{1000}) \: \eta_{\rm w0} + ({S_{\rm A}}/{1000}) \: \eta_{\rm s0}$,
which can be rewritten as 
$[\:\eta_{\rm w0}\:] + [\:(\eta_{\rm s0} - \eta_{\rm w0}) \: ({S_{\rm A}}/{1000})\:]$. 
This explains the natural splitting of this sum 
in the two TEOS10's Tables~\ref{Table_TEOS10_sigma_W_x_y_z}
and \ref{Table_TEOS10_sigma_S_x_y_z} 
(for $\eta^{\rm W}_{\rm Fei03}$ and $\eta^{\rm S}_{\rm Fei08}$) 
into the two bracketed 
`\,pure-water\,' $[\:\eta_{\rm w0}\:]$ 
and `\,salinity\,' 
$[\:(\eta_{\rm s0} - \eta_{\rm w0}) \: ({S_{\rm A}}/{1000})\:]$
parts, respectively.
Other linear and non-linear variables terms generated  
by the three lines of (\ref{eq_FH95_4_9}) are similarly
cast into the `\,pure-water\,' and `\,salinity\,' parts of
$g$, and thus of $\eta$. In particular, the second line of 
(\ref{eq_FH95_4_9}) explains the term $x^2\:\ln(x)$ with
$x^2=S_{\rm A}/40.188\,617=24.882\,67\:C$.

Therefore, the reference values of the entropies do not impact 
the pure-water part of the seawater entropy, since $[\:\eta_{\rm w0}\:]$
is a mere constant with no physical meaning (null differential).
Differently, the linear first-order salinity term 
must have a physical impact on the computation of 
the seawater entropy. 
Indeed, its differential 
$(\eta_{\rm s0} - \eta_{\rm w0}) \: (d{S_{\rm A}}/{1000})$
is a priori different from zero and depends on 
the difference of the two reference values.  
Thus, it depends on both $\eta_{\rm s0}$ and $\eta_{\rm w0}$,
which must be determined in order to compute the absolute value 
of the seawater entropy $\eta$ in 
(\ref{eq_TEOS10_eta_x_y_z}) from 
Tables~\ref{Table_TEOS10_sigma_W_x_y_z} and 
\ref{Table_TEOS10_sigma_S_x_y_z}. 

The first choice retained in pre-TEOS10 formulations was 
to arbitrarily set $\eta_{\rm w0}=0$ in order to cancel the 
pure-water entropy at the triple-point conditions: 
$\eta^{\rm W}_{\rm Fei03}(T_{\rm t},p_{\rm t})=0$
\citep[][Eq.~G.2, p.~155]{Feistel_TEOS_manual_2010}.
In fact, in the present version of TEOS10 software,  
the term $\eta_{\rm w0}$ is set to 
$-g_{010}/40 \approx -\,5.905\,6/40 \approx 0.147\,64$
instead, and 
$(\eta_{\rm s0} - \eta_{\rm w0})$ is set to
$-g_{210}/40 \approx -\,168.072\,4/40 \approx 4.201\,8$  
in order to arrive at the cancellation of the 
whole seawater entropy $\eta(S_{\rm SO},t_{\rm SO},p_{\rm SO})$  
for the arbitrary standard conditions 
$S_{\rm SO}=35.165\,04$~g~kg${}^{-1}$, $t_{\rm SO}=0{}^{\circ}$C, 
and $p_{\rm SO}=0$~dbar
\citep[see the explanations p.~155 and Eq.~2.6.6, 
p.~17 of][]{Feistel_TEOS_manual_2010}.

\begin{table*}[hbt] 
\caption{\it\small The ``\,pure-liquid-water\,'' part of the 
seawater entropy function as computed from the TEOS10 software 
(see the $g_{ijk}$ coefficients in Appendix~G, p.~155),
where $y = t / (40{}^{\circ}$C$\,)$ and 
$z = p / (100$~MPa$\,)$ are two dimensionless variables.} 
\vspace*{-5mm}
\begin{align}
 \overbrace{
 \eta^{\rm W}_{\rm Fei03}\:(y,z) 
 }^{\mbox{``\,pure-Water\,''}}
 & \; = \;
 \overbrace{
    \left\{ \eta_{\rm w0}  \:\right\} 
 }^{\mbox{``\,???\,''}}
 \; + \: 
    \left(\frac{1}{40}\right) \:
  \overbrace{(
  {
  -\,5.90578347909402\;
  )}}^{\displaystyle (-g_{010})} \: 
 \; + \: 
    \left(\frac{1}{40}\right)  
 \,[ \:\; 
 270.983805184062    \: z 
 \nonumber \\
 &  \quad
 -776.153611613101    \: z^2
+ 196.51255088122     \: z^3
-  28.9796526294175   \: z^4
 \nonumber \\
&  \quad
    +2.13290083518327 \: z^5
+24715.571866078      \: y   
 -2910.0729080936     \: y \: z 
 \nonumber \\
&  \quad
 +1513.116771538718   \: y \: z^2
  -546.959324647056   \: y \: z^3
  +111.1208127634436  \: y \: z^4
 \nonumber \\
&  \quad
    -8.68841343834394 \: y \: z^5
-2210.2236124548363   \: y^2 
+2017.52334943521     \: y^2 \: z 
 \nonumber \\
&  \quad
-1498.081172457456    \: y^2 \: z^2 
 +718.6359919632359    \: y^2 \: z^3 
 -146.4037555781616    \: y^2 \: z^4 
 \nonumber \\
&  \quad
   +4.9892131862671505 \: y^2 \: z^5 
 +592.743745734632     \: y^3   
-1591.873781627888     \: y^3 \: z 
 \nonumber \\
&  \quad
+1207.261522487504     \: y^3 \: z^2
 -608.785486935364 \: y^3 \: z^3
+105.4993508931208 \: y^3 \: z^4
 \nonumber \\
&  \quad
-290.12956292128547 \: y^4
+973.091553087975 \: y^4 \: z 
-602.603274510125 \: y^4 \: z^2
 \nonumber \\
&  \quad
+276.361526170076 \: y^4 \: z^3
 -32.40953340386105 \: y^4 \: z^4
 +113.90630790850321  \: y^5   
 \nonumber \\
&  \quad
-21.35571525415769 \: y^6  
+67.41756835751434 \: y^6 \: z
-381.06836198507096   \: y^5 \: z 
 \nonumber \\
&  \quad
+133.7383902842754   \: y^5 \: z^2
-49.023632509086724  \: y^5 \: z^3 \:] 
\nonumber \: .  
\end{align}
\label{Table_TEOS10_sigma_W_x_y_z}
\vspace*{-8mm} \\
---------------------------------------------------------------------------------------------------------------------------------
\end{table*}

\begin{table*}[hbt] 
\caption{\it\small The ``\,saline\,'' (sea-salts) part of the seawater entropy 
function as computed from the TEOS10 software
(see the $g_{ijk}$ coefficients in Appendix~H, p.~156),
with the dimensionless concentrations $C=X_{\rm A}=S_{\rm A}/1000$ and $X_{\rm SO}=S_{\rm SO}/1000$ 
and the dimensionless variables $x^2 = S_{\rm A}/(40.188\,617\,$g~kg${}^{-1})$, $y = t / (40{}^{\circ}$C$\,)$ 
and $z = p / (100$~MPa$\,)$, where $S_{\rm A}$ is the absolute salinity in units of g~kg${}^{-1}$ 
and $S_{\rm SO}=35.165\,04$~g~kg${}^{-1}$ is a standard seawater salinity (see Table~D.4, p.~145).}
\vspace*{-5mm}
\begin{align}
 \overbrace{
 \eta^{\rm S}_{\rm Fei08}\:(x,y,z) 
 }^{\mbox{``\,Salinity\,''}}
 &\; = \;
 \overbrace{
 \left\{\: 
    \eta_{\rm s0} - \eta_{\rm w0} \: 
 \:\right\} 
 }^{\mbox{``\,???\,''}}
 \;\times\; \left( \frac{S_{\rm A} - S_{\rm SO}}{1000} \right)  
 \; + \: 
 \left(\frac{1}{40}\right)  
 \,[\:\;
  \overbrace{({
    -\,168.072408311545 }
  )}^{\displaystyle 
       (-g_{210})} \: x^{2} 
 \nonumber \\
&  \quad
    -851.226734946706 \: \ln(x)\: x^2 
    -729.116529735046 \: x^2 \: z
    +343.956902961561 \: x^2 \: z^2
 \nonumber \\
&  \quad
    -124.687671116248 \: x^2 \: z^3 
    +31.656964386073  \: x^2 \: z^4
    -7.04658803315449 \: x^2 \: z^5
 \nonumber \\
&  \quad
    +493.407510141682 \: x^3 
    -543.835333000098 \: x^4 
    +196.028306689776 \: x^5
 \nonumber \\
&  \quad
     -36.7571622995805 \: x^6  
    +137.1145018408982 \: x^4 \: y
   -148.10030845687618 \: x^4 \: y^2
 \nonumber \\
&  \quad 
    +68.5590309679152  \: x^4 \: y^3  
    -12.4848504784754  \: x^4 \: y^4  
    +22.6683558512829  \: x^4 \: z
 \nonumber \\
&  \quad 
   +175.292041186547   \: x^3 \: z
    -83.1923927801819  \: x^3 \: z^2
    +29.483064349429   \: x^3 \: z^3 
 \nonumber \\
&  \quad 
    +86.1329351956084  \: x^3 \: y
   -766.116132004952   \: x^3 \: y \: z
   +108.30162043765552 \: x^2 \: y^4  
 \nonumber \\
&  \quad 
    -51.2796974779828  \: x^3 \: y \: z^3 
    +30.068211258562   \: x^3 \: y^2
  +1380.9597954037708  \: x^3 \: y^2 \: z
 \nonumber \\
&  \quad 
     -3.50240264723578 \: x^3 \: y^3  
   -938.26075044542    \: x^3 \: y^3 \: z 
  -1760.062705994408   \: x^2 \: y
 \nonumber \\
&  \quad  
   +675.802947790203   \: x^2 \: y^2
   -365.7041791005036  \: x^2 \: y^3
   +108.3834525034224  \: x^3 \: y \: z^2
 \nonumber \\
&  \quad 
    -12.78101825083098 \: x^2 \: y^5  
  +1190.914967948748   \: x^2 \: y^3 \: z 
 \nonumber \\
&  \quad 
   -298.904564555024   \: x^2 \: y^3 \: z^2
   +145.9491676006352  \: x^2 \: y^3 \: z^3
  \nonumber \\
&  \quad 
  -2082.7344423998043  \: x^2 \: y^2 \: z
   +614.668925894709   \: x^2 \: y^2 \: z^2
  \nonumber \\
&  \quad 
   -340.685093521782   \: x^2 \: y^2 \: z^3
    +33.3848202979239  \: x^2 \: y^2 \: z^4
 \nonumber \\
&  \quad
  +1721.528607567954   \: x^2 \: y  \: z
   -674.819060538734   \: x^2 \: y  \: z^2
   +356.629112415276   \: x^2 \: y \: z^3 
 \nonumber \\
&  \quad
   -88.4080716616      \: x^2 \: y   \: z^4
   +15.84003094423364  \: x^2 \: y   \: z^5 \:] 
\nonumber \: . 
\end{align}
\label{Table_TEOS10_sigma_S_x_y_z}
\vspace*{-8mm} \\
---------------------------------------------------------------------------------------------------------------------------------
\end{table*}

\begin{table*}[hbt] 
\caption{\it\small The molar mass ($M$, in~g~mol\,${}^{-1}$),
 mole fraction ($X$, in~mol~mol\,${}^{-1}$) 
 and absolute entropies ($S$ , in J~K${}^{-1}$~mol${}^{-1}$,
 `relative' to $\mbox{S(H${}^{+}$)} = 0 $),   
 computed at $25\,{}^{\circ}\,$C and $0.1$~MPa,  
 for the pure liquid water (H${}_{2}$O)${}_{\rm liq.}$ 
 and the main seasalts (cations and anions).
Values of $M${(M82)} are from \citet[][Table IV, p.~428]{Millero_1982};  
those of $X${(M83)} are from M83 (Table~X, p.~35) 
with about half the values of $X${(M83)} normalized to a sum of $1$; 
those of $M$(TEOS10) and $X$(TEOS10) are from 
\citet[][Table~D.3, p.~137]{Feistel_TEOS_manual_2010}, 
with the molar fractions obtained from the rounded values of
\citet[][Table~3, p.~60]{Millero_al_2008}.
The absolute entropies 
``M83'' (from Table~X, p.~35) 
are the same as those in ML76 (Table~32, p.~1071) 
and are in fact those previously listed in ``LR61'' 
(Table~25.7, p.~400-401). 
The more recent entropies ``G92'' are from 
\citet[][``NEA-TDB'' for ``Nuclear Energy Agency and 
Thermodynamic Data Bank'', see the 
``Selected auxiliary data'' p.~64-83]{Grenthe_al_NEA_TDB_1992},
which are still retained in 
\citet{Grenthe_al_NEA_TDB_2020}.
The numbers in parentheses refer to the uncertainty in 
the corresponding last digits, with, for instance, 
$96.062\,6(50)$ meaning $96.062\,6\: \pm 0.005\,0$.}
\vspace*{2mm}
\centering
\begin{tabular}{cccccccccc}
\hline 
 & $M${(M82)}
 & $X${(M83)}
 & $S${(LR61/M83)}
 & $M${(TEOS10)}
 & $X${(TEOS10)}
 & $S${(G92)}
\\ \hline  
(H${}_{2}$O)${}_{\rm liq.}$
& 
& 
& $70.00$
& $18.015\,268$
& 
& $69.95(3)$
\\ \hline  
{(H${}^{+}$})
& 
& 
& $(0.0)$ 
& 
&
& $(0.0)$ 
\\   
{Na${}^{+}$}
& {$22.989\,8$}  
& {$0.418\,212\,1$}
& $60.2$ 
& {$22.989\,769\,28(2)$}
& {$0.418\,807\,1$}
& $58.45(15)$
\\   
{Mg${}^{2+}$}
& {$24.305$}  
& {$0.047\,558\,3$}
& $-118$ 
& {$24.305\,0(6)$}
& {$0.047\,167\,8$}
& $-137(4)$
\\   
{Ca${}^{2+}$}
& {$40.08$}  
& {$0.009\,172\,6$}
& $-55.2$ 
& {$40.078(4)$}
& {$0.009\,182\,3$}
& $-56.2(10)$ 
\\   
{K${}^{+}$}
& {$39.102$}  
& {$0.009\,112\,6$}
& $102.5$ 
& {$39.098\,3(1)$}
& {$0.009\,115\,9$}
& $101.2(2)$
\\   
Sr${}^{2+}$
& $87.62$ 
& $0.000\,080\,0$
& $-39.3$ 
& $87.62(1)$
& $0.000\,081\,0$
& $-31.5(20)$
\\   
{Cl${}^{-}$}
& {$35.453$}  
& {$0.487\,541\,5$}
& $55.2$ 
& {$35.453(2)$} 
& {$0.487\,483\,9$}
& $56.6(2)$ 
\\   
{SO${}_4^{2-}$}
& {$96.057\,6$}
& {$0.025\,217\,1$}
& $17.2$ 
& {$96.062\,6(50)$}
& {$0.025\,215\,2$}
& $18.5(4)$
\\   
{HCO${}_3^{-}$}
& {$61.017\,2$}
& {$0.001\,725\,5$}
& $95.0$ 
& {$61.016\,84(96)$}
& {$0.001\,534\,0$}
& $98.4(5)$
\\   
{Br${}^{-}$}
& {$79.904$}
& {$0.000\,755\,2$}
& $80.8$ 
& {$79.904(1)$}
& {$0.000\,752\,0$}
& $82.55(20)$
\\   
CO${}_3^{2-}$
& $60.009\,2$
& $0.000\,200\,0$
& $-51.3$ 
& $60.008\,9(10)$
& $0.000\,213\,4$
& $-50(1)$
\\   
B(OH)${}_4^{-}$
& $78.839\,6$
& $0.000\,075\,0$
& $(\approx 100)$ 
& $78.840\,4(70)$
& $0.000\,090\,0$
& $(102.5)$
\\   
F${}^{-}$
& $18.998\,4$
& $0.000\,050\,0$
& $-9.6$ 
& $18.998\,403\,2(5)$
& $0.000\,061\,0$
& $-13.8(8)$
\\   
OH${}^{-}$
& $(-)$
& $(-)$
& $-10.5$ 
& $17.007\,33(7)$
& $0.000\,007\,1$
& $-10.9(2)$
\\   
B(OH)${}^{\rm aq.}_{3}$
& $61.832\,2$
& $0.000\,300\,1$
& $(\approx 50)$ 
& $61.833\,0(70)$
& $0.000\,280\,7$
& $162.4(6)$
\\  
CO${}^{\rm aq.}_{2}$
& $(-)$
& $(-)$
& $(-)$ 
& $44.009\,5(9)$
& $0.000\,008\,6$
& $119.36(60)$
\\ \hline   
Mean
& $31.405$
& $1.000\,000\,0$
& $47.52$ 
& $31.404(1)$
& $1.000\,000\,0$
& $46.74(40)$
\\ \hline 
\end{tabular}
\label{Table_TEOS10_M_X_S}
\end{table*}




 \subsection{A need to update the seawater entropy}
\label{subsection_need_update}
\vspace*{-2mm}

The seawater entropy can be written as
$\eta(x,y,z) = \eta_{\rm w0} 
+ (\eta_{\rm s0} - \eta_{\rm w0}) \: ({S_{\rm A}}/{1000})
+ (...)$, 
where $(...)$ represents the non-linear higher-order 
terms in $x^2$, $y$ or $z$. 
Therefore, the first-order terms 
of the vertical gradient $\partial \eta / \partial z$ 
and the turbulent flux $\overline{w'\:\eta'}$ 
are the product of the measurable quantities 
$\partial S_{\rm A} / \partial z$ and $\overline{w'\:S'_{\rm A}}$ by the 
difference in reference entropy $(\eta_{\rm s0} - \eta_{\rm w0})$.
Then, since the entropy is a thermodynamic state function, 
the difference of $\eta$ between two points, and thus its gradients 
and turbulent fluxes, cannot be arbitrary and cannot be either 
positive, null, or negative, depending on the values and signs of
$\eta_{\rm s0} - \eta_{\rm w0}$.
This means that it is not possible to modify or arbitrarily set 
$\eta_{\rm s0}$ and/or $\eta_{\rm w0}$ to this or that value 
unless modifying the entropy budget 
($d\eta/dt=...$), and thus the second law of thermodynamics.

It may be decided not to bother with entropy calculations, 
since the reference entropy values have no impact on the other 
thermodynamic variables (specific volumes, heat capacities, 
expansion coefficients, sound speeds, osmotic pressure, etc.) 
and only impact the entropy, Helmholtz 
free energy, and Gibbs free enthalpy functions. 
But in this case, we have to go right to the end of the 
consequences and refrain from calculating values of 
these three functions, leaving these three quantities 
undetermined to within a function of salinity. 

Differently, one of the aims of TEOS10, as 
Millero 
reminded us, is to provide precise values for these three 
functions.
So there is no choice if you want to calculate these three 
functions: you have to leave no room for arbitrariness and 
trust the recommendations of general thermodynamics.
Here lies the strong motivation for using non-arbitrary values
for not only the difference in reference entropies 
$\eta_{\rm s0} - \eta_{\rm w0}$, 
but in fact for both $\eta_{\rm s0}$ and $\eta_{\rm w0}$, 
and thus to use the absolute, third-law values of them defined 
in thermodynamics.

In addition, although the formulations of ML76 and M83 
brought a clear theoretical improvement in taking into account 
the absolute reference entropies of liquid water and 
sea salts, the analytical and numerical formulation seems somewhat 
anomalous (see subsection~\ref{subsection_abs_entropy_Millero}).
As a consequence, Millero's calculations had to be 
restated using more recent data and methods.

 \subsection{The absolute entropy of pure liquid water} 
\label{subsection_abs_entropies_water}
\vspace*{-2mm}

I recall in the second line of Table~\ref{Table_TEOS10_M_X_S} 
that the value of the absolute entropy of (pure) liquid water is well-known 
since at least LR61 (Table 12.3, p.~137), 
who published values corresponding to
\begin{align}
\eta_{\rm w0/LR61}\mbox{($25{}^{\circ}$C)} & \approx 16.73 \times 4.184 
\nonumber \: , \\
\eta_{\rm w0/LR61}\mbox{($25{}^{\circ}$C)} & 
\approx 70.00\mbox{~J~K${}^{\,-1}$~mol${}^{\,-1}$} 
\label{Eq_ML76_M83_eta_w0_mol} \; , \\
\eta_{\rm w0/LR61}\mbox{($25{}^{\circ}$C)} & \approx \frac{70.00}{0.018\,015} 
\nonumber \: , \\
\eta_{\rm w0/LR61}\mbox{($25{}^{\circ}$C)} &  
\approx 3885.65\mbox{~J~K${}^{\,-1}$~kg${}^{\,-1}$}
\label{Eq_ML76_M83_eta_w0_kg} \; . 
\end{align}
These values retained in ML76 and M83 correspond to the molar mass  
$M_{\rm w}=0.018\,015$~kg${}^{\,-1}$~mol${}^{\,-1}$.
The use of a mean specific heat of  
$c_{\rm w} \approx 4218$~J~K${}^{\,-1}$~kg${}^{\,-1}$ leads
to the value $3516$~J~K${}^{\,-1}$~kg${}^{\,-1}$
at $0{}^{\circ}$C$\,$\footnote{$\:$The corresponding correction term  
$4218 \times \ln(273.15/298.15) \approx -369.39$~J~K${}^{\,-1}$~kg${}^{\,-1}$ 
is about $-2$ units smaller than the similar value 
$-367.36$~J~K${}^{\,-1}$~kg${}^{\,-1}$ 
computed with the TEOS10's (GSW) 
subroutine ({\tt{liq\_entropy\_si}}), 
leading to a smaller TEOS10's mean value 
$\overline{c_{\rm w}} \approx 4194.6$~J~K${}^{\,-1}$~kg${}^{\,-1}$ 
between $298.15$~K and $273.15$~K 
and to a larger LR61 and ML76 pure-liquid water entropy 
of about $3518$~J~K${}^{\,-1}$~kg${}^{\,-1}$ at $273.15$~K.}.
This value, represented by the purple disk 
in Fig.~\ref{Fig_Entropy_H2O_stat_calor}(b),
agrees with the value $3522 \pm 12$~J~K${}^{-1}$~kg${}^{-1}$ 
from \citet{Giauque_Stout_36} recalled by 
\citet[][p.~104]{Feistel_Wagner_2005} in the improved IAPWS-95 paper. 
It also agrees with the other values cited by 
\citet[][p.~104]{Feistel_Wagner_2005}: 
$69.96 \pm 0.03$~J~K${}^{\,-1}$~mol${}^{\,-1}$ 
\citep[from][]{Cox_al_CODATA_1989} and 
$70.07 \pm 0.22$~J~K${}^{\,-1}$~mol${}^{\,-1}$ 
(computed by Feistel and Wagner), 
and is the same as the updated value of 
$3516 \pm 2$~J~K${}^{\,-1}$~kg${}^{\,-1}$
next considered by \citet[][p.~1032]{Feistel_Wagner_2006}. 
Moreover this value also includes the residual entropy at $0$~K of 
$3.407 \pm 0.001$~J~K${}^{-1}$~mol${}^{-1}$ or 
$189.13 \pm 0.05$~J~K${}^{-1}$~kg${}^{-1}$ 
\citep{Pauling1935,Nagle1966}.

I will retain for this study the smaller value 
\begin{align}
 \eta_{\rm w0}\mbox{($0{}^{\circ}$C)} & \: \approx 3513.4 
 \pm 1.7~\mbox{J~K${}^{\,-1}$~kg${}^{\,-1}$}
 \label{eq_eta_w0_NEA_TDB_1992} \: 
\end{align}
corresponding to  
$69.95 \pm 0.03$~J~K${}^{\,-1}$~mol${}^{\,-1}$ at $25{}^{\circ}$C 
published in the NEA-TDB Tables by
\citet[][p.~64]{Grenthe_al_NEA_TDB_1992} 
and  
\citet[][p.~149]{Grenthe_al_NEA_TDB_2020} 
recalled in Table~\ref{Table_TEOS10_M_X_S}.  
This value is also compatible with the one  
($69.950 \pm 0.079$~J~K${}^{\,-1}$~mol${}^{\,-1}$) 
published in the NIST-JANAF4 Tables by
\citet[][p.~1323]{Chase_1998}, 
which is also to $69.9422$~J~K${}^{\,-1}$~mol${}^{\,-1}$ 
in Table~B.6 by \citet[][p.~889]{Schmidt_Thermodynamics_Engineers_Springer_2022} 
and also to $69.91$~J~K${}^{\,-1}$~mol${}^{\,-1}$ 
in the even more recent Tables by \citet[][p.~893]{Atkins_Paula_Keeler_2023}.

 \subsection{The absolute entropy of sea-salts at  $25{}^{\circ}$C}
\label{subsection_abs_entropies_seasalts_25C}
\vspace*{-2mm}

The mean absolute entropy for sea-salt aqueous ions 
has already been computed 
by  ML76 (Table~32, p.~1071) and M83 (Table~X, p.~35)
from the individual values $S$(LR61/M83) previously published 
by LR61 (Table~25.7, p.~400-401) 
and recalled in Table~\ref{Table_TEOS10_M_X_S}.
The resulting value was
\begin{align}
\eta_{\rm s0/M83}\mbox{($25{}^{\circ}$C)} & 
\approx \frac{95.01}{1.999\,44} 
\nonumber \; , \\
\eta_{\rm s0/M83}\mbox{($25{}^{\circ}$C)} &  
\approx 47.52\mbox{~J~K${}^{\,-1}$~mol${}^{\,-1}$} 
\label{Eq_ML76_M83_eta_s0_mol} \; , 
\end{align}
and
\begin{align}
\eta_{\rm s0/M83}\mbox{($25{}^{\circ}$C)} & 
\approx \frac{95.01}{0.062\,793} 
\nonumber \; , \\
\eta_{\rm s0/M83}\mbox{($25{}^{\circ}$C)} & 
\approx 1513.1\mbox{~J~K${}^{\,-1}$~kg${}^{\,-1}$}
\label{Eq_ML76_M83_eta_s0_kg} \; . 
\end{align}
The factor $1.999\,44 \approx 2$ (see p.~87 in the SM) 
was due to a definition of 
molal concentrations and molar entropies published by 
ML76 (text p.~1070, Table~32 p.~1071 and Appendix p.~1074) and 
M83 (text and Table-X p.~35) that are 
different from those shown in the columns $X$(TEOS10) and $M$(TEOS10) 
in Table~\ref{Table_TEOS10_M_X_S} and published 
by \citet[][Tables~3 and 4, p.~60 and p.~62]{Millero_al_2008} and
\citet[][Table~D.3, p.~137]{Feistel_TEOS_manual_2010}. 

Therefore, the mean sea-salt molar mass previously derived by ML76 
($M_2$ in Eq.~10 in the Appendix, p.~1074) roughly corresponds to 
$<\!A\!>\:=62.793 /1.999\,44 \approx 31.405$~g~mol${}^{\,-1}$.
More precisely, more relevant definitions were suggested 
by \citet[][Eq.~88, p.~427 and Table~IV, p.~428]{Millero_1982}, 
with the total mole written as 
$n_{\rm T}=n_{\rm B} + (1/2) \sum_i n_i$
and leading to a sum 
$N_{\rm B} + (1/2) \sum_i N_i = 1.000\,000\,5$ 
very close to $1$ and to 
$M_{\rm T} = N_{\rm B}\:M_{\rm B} + (1/2) \sum_i N_i\:M_i 
\approx 31.3677$~g~mol${}^{-1}$  
indeed indicating double values for all the ions 
$N_i \neq N_{\rm B}$ (i.e., except the non-ionic Boric term). 
However, these more relevant 1982 definitions were not 
considered in ML76 and will not be retained in M83.

Note that the sea-salt aqueous ions listed 
in Table~\ref{Table_TEOS10_M_X_S} are relative 
to $\overline{S}^{\,\circ}_{\ch{H+}}=0$, with, 
however, no impact of a non-zero value for
$\overline{S}^{\,\circ}_{\ch{H+}}$ on the mean value 
for the sea-salt entropy. 
Indeed, a non-zero value $\Delta S_0$ for $\overline{S}^{\,\circ}_{\ch{H+}}$ 
would lead to a change in the average sea salt entropy corresponding 
to the weighted sum $\sum_j (Z_j \: \Delta S_0) \: X_j 
= (\sum_j Z_j \: X_j) \: \Delta S_0$, 
which depends on the valence (ionic charges) $Z_j$ 
and the mole fractions $X_j$.
Fortunately, this sum is zero in the Millero (ML76 and M83) 
and TEOS10 set-ups thanks to the neutrality property 
$\sum_i Z_j \: X_j = 0$ \citep[see the column for 
$Z_j \times X_j$ in Table~D.3 of][p.~144]{Feistel_TEOS_manual_2010}. 

Moreover, Millero's articles suffer from the fact that they do not take into 
account the two species OH${}^{-}$ and CO${}^{aq.}_{2}$ considered 
in the more recent studies and TEOS10.

Due to all these inaccuracies and outdated references in the papers by ML76 
and M83, it was desirable to use updated and more recent values for 
the sea-salts entropies and concentrations, like the TEOS10's sea-salts values 
listed in \citet[][Table~D.3, p.~144]{Feistel_TEOS_manual_2010}.  
It was also necessary to use updated values for 
the molar absolute sea-salt entropies $\Delta_r S^{\circ}_m$, 
like those listed in the ``Selected auxiliary data'' 
in the NEA-TDB Tables~IV.1 and IV.2 of 
\citet[][S(G92), p.~64-83]{Grenthe_al_NEA_TDB_1992} 
and retained in \citet{Grenthe_al_NEA_TDB_2020}, 
including uncertainty intervals.

\begin{table*}[hbt] 
\centering
\caption{\it\small Values of the molar specific heat at constant pressure 
$\overline{C}^{\,\circ}_p$ at $25{}^{\circ}$C (in J~K${}^{-1}$~mol${}^{-1}$)
for several of the major sea-salt cations and anions, and based on
$\overline{C}^{\,\circ}_p \mbox{(H${}^{+}$)} = 0$: 
``LR61'' (Table~25-7, p.~400-401)
\citet[][Table~25-7, p.~400-401]{Lewis_Randall_1961}; 
``NBS82'' from \citet[][Tables p.~2-47, 2-50, 2-57, 2-83, 
         2-260, 2-267, 2-299, 2-328]{Wagman_al_NBS_1982}; 
 ``CM96'' from \citet[][Table~2, 
    p.~1290]{Criss_Millero_Heat_Capacities_Electrolyte_1996}; 
 ``HH96'' from 
\citet[][Table~5, p.~647]{Hepler_Hovey_heat_capacities_electrolytes_1996}.
Conversions from the old unit (cal~K${}^{-1}$~mol${}^{-1}$) of 
``LR61'' are made with $1$~cal$\:=4.184$~J.
The values ``M12''   
from \citet[][Table~1.2, p.~12]{Marcus_Ions_properties_2012}
were with $\overline{C}^{\,\circ}_p \mbox{(H${}^{+}$)} 
= -71$~J~K${}^{-1}$~mol${}^{-1}$ and have been transformed 
for the reference $\overline{C}^{\,\circ}_p \mbox{(H${}^{+}$)} = 0$
by taking into account the valences of
 ions $Z$ ($-2$, $-1$, $+1$, $+2$). 
The molar concentrations $X$ (in mol~mol${}^{-1}$) are
adapted from \citet[][Table~D.3, p.~137]{Feistel_TEOS_manual_2010}
recalled in the previous Table~\ref{Table_TEOS10_M_X_S}.
The rescaled values lead to a sum of 1 for either the
case $X(2)$ where only two major species (Na${}^{+}$ and Cl${}^{-}$)
are available, or the case $X(4)$ where four major species 
(Na${}^{+}$, Mg${}^{2+}$, Cl${}^{-}$ and SO${}_4^{2-}$)
are available, or the case $X(8)$ where all eight species 
are available.}
\vspace*{2mm}
\begin{tabular}{cccccccccc}
\hline 
 $\overline{C}^{\,\circ}_p$
 & {Na${}^{+}$}
 & {Mg${}^{2+}$}
 & {Ca${}^{2+}$}
 & {K${}^{+}$}
 & {Cl${}^{-}$}
 & {SO${}_4^{2-}$}
 & {HCO${}_3^{-}$}
 & {Br${}^{-}$}
 & Mean
\\ \hline 
   $X(2)$
 &   {$0.500$}
 &   
 &   
 &   
 &   {$0.500$}
 &   
 &  
 &   
 & {$1.000$}
\\ 
  LR61 
 & {$33.1$}
 & {($+16.7$)}
 & {($-37.7$)}
 & {($9.6$)}
 & {$-126$}
 & {($-276$)}
 & {(...)}
 & {($-130$)}
 & {$-\,46.5$}
\\ 
  NBS82
 & {$46.4$}
 & {(...)}
 & {(...)}
 & {($21.8$)}
 & {$-136.4$}
 & {($-293$)}
 & {(...)}
 & {($-141.8$)}
 & {$-45.0$}
\\ 
  CM96
 & {$43.01$}
 & {(...)}
 & {(...)}
 & {($12.47$)}
 & {$-126.32$}
 & {(...)}
 & {($-53.33$)}
 & {($-131.27$)}
 & {$-41.7$}
\\ 
  HH96 
 & {$42$}
 & {($-16$)}
 & {($-27$)}
 & {($12$)}
 & {$-126$}
 & {($-276$)}
 & {($-52$)}
 & {($-132$)}
 & {$-42.0$}
\\  
  M12 
 & {$43$}
 & {($-16$)}
 & {($-27$)}
 & {($13$)}
 & {$-127$}
 & {($-280$)}
 & {($-53$)}
 & {($-131$)}
 & {$-42.0$}
\\ \hline 
   $X(4)$ 
 & {$0.427\,93$}
 & {$0.048\,20$}
 &  
 &   
 & {$0.498\,10$}
 & {$0.025\,77$}
 &   
 &  
 & {$1.000$}
\\ 
  LR61 
 & {$33.1$}
 & {$+16.7$}
 & {($-37.7$)}
 & {($9.6$)}
 & {$-126$}
 & {$-276$}
 & {(...)}
 & {($-130$)}
 & {$-\,54.9$}
\\ 
  HH96 
 & {$42$}
 & {$-16$}
 & {($-27$)}
 & {($12$)}
 & {$-126$}
 & {$-276$}
 & {($-52$)}
 & {($-132$)}
 & {$-52.7$}
\\  
  M12 
 & {$43$}
 & {$-16$}
 & {($-27$)}
 & {($13$)}
 & {$-127$}
 & {$-280$}
 & {($-53$)}
 & {($-131$)}
 & {$-52.8$}
\\ \hline 
   $X(8)$ 
 & {$0.419\,12$}
 & {$0.047\,20$}
 & {$0.009\,19$}
 & {$0.009\,13$}
 & {$0.487\,84$}
 & {$0.025\,24$}
 & {$0.001\,53$}
 & {$0.000\,75$}
 & {$1.000$}
\\ 
  HH96 
 & {$42$}
 & {$-16$}
 & {$-27$}
 & {$12$}
 & {$-126$}
 & {$-276$}
 & {$-52$}
 & {$-132$}
 & {$-51.9$}
\\  
  M12 
 & {$43$}
 & {$-16$}
 & {$-27$}
 & {$13$}
 & {$-127$}
 & {$-280$}
 & {$-53$}
 & {$-131$}
 & {$-52.1$}
\\ \hline  
\end{tabular}
\label{Table_S_water_salts}
\end{table*}

The corresponding updated mean values for $M$(TEOS10) and 
$S$(G92) shown in Table~\ref{Table_TEOS10_M_X_S}
are 
\begin{align}
M_{\rm s} & \:\approx\; 31.404 \pm 0.001~\mbox{g~mol${}^{\,-1}$}
\label{Eq_Ms} \: , 
\end{align}
\begin{align}
\eta_{\rm s0}\mbox{($25{}^{\circ}$C)} & \:\approx\;
  46.74 \pm 0.40~\mbox{J~K${}^{\,-1}$~mol${}^{\,-1}$}
\nonumber 
\: , \\
\eta_{\rm s0}\mbox{($25{}^{\circ}$C)} & \:\approx\;
  1488.3 \pm 13~\mbox{J~K${}^{\,-1}$~kg${}^{\,-1}$}
\label{Eq_eta_s_25C_kg} \: ,
\end{align}
which is only $24.8$~J~K${}^{\,-1}$~kg${}^{\,-1}$ and 
$1.6$~\% smaller than the Millero's value.
I have also computed (not shown) similar sea-salt 
entropies at $25{}^{\circ}$C from other datasets: 
 $47.33 \pm 2$~J~K${}^{\,-1}$~mol${}^{\,-1}$
                  from \citet[][Table~III, p.~84]{Latimer_al_1934},
 $47.04$         from \citet[][Table~II, p.~1831]{Latimer_al_1938},
 $46.75 \pm 0.4$ from \citet[][USG-1452]{Robie_al_1978}, and
 $46.93$         from \citet[][NBS-82]{Wagman_al_NBS_1982}.

 \subsection{The absolute entropy of sea salts at $0{}^{\circ}$C}
\label{subsection_abs_entropies_seasalts_0C}
\vspace*{-2mm}

In order to compute the mean reference sea-salt entropy 
at $0{}^{\circ}$C by using the relationship 
$S(273.15)=S(295.15)+\overline{C}^{\,\circ}_{p}\:\ln(273.15/298.15)$,
it is needed to know the mean specific heats of sea salts at 
constant pressure $\overline{C}^{\,\circ}_{p}$, for instance
by averaging the individual values for each sea salt recalled 
in Table~\ref{Table_S_water_salts} from several datasets, 
where individual values are not available for all species.
For this reason, three kinds of averaging are made:
first with the two main species  
$0.5$\% of {Na${}^{+}$} and $0.5$\% of {Cl${}^{-}$}
and the concentrations $X(2)$;
then for the four main species
Na${}^{+}$, Mg${}^{2+}$, Cl${}^{-}$, and SO${}_4^{2-}$ 
(for LR61, HH96 and M12)
and the concentrations $X(4)$; 
and then for the whole set
of the eight species (for HH96 and M12 only)
and the concentrations $X(8)$.

When available, the order of magnitude and the signs of 
$\overline{C}^{\,\circ}_p$ are similar for the $6$ species 
Na${}^{+}$, K${}^{+}$, Cl${}^{-}$, SO${}_4^{2-}$, 
HCO${}_3^{-}$ and Br${}^{-}$.
The discrepancies are larger for {Ca${}^{2+}$} and {Mg${}^{2+}$},  
with even not the same signs for {Mg${}^{2+}$}. 
However, the value for {Mg${}^{2+}$} computed in a recent paper by 
\citet[][Table~7, p.~H]{Caro_al_Cp_Mg2plus_2020} is negative ($-155$) 
and close to the value $-16 - 2 \times 71 = -158$ 
published in both HH96 and M12. 
For this reason the old positive values of LR61  
for {Mg${}^{2+}$} might be inaccurate. 

The mean values computed with $0.5$\% of {Na${}^{+}$} and $0.5$\% of 
{Cl${}^{-}$} are close to $-42$ to $-46$ units.
The mean values computed with the $4$ major species 
Na${}^{+}$, Mg${}^{2+}$, Cl${}^{-}$ and SO${}_4^{2-}$
are more negative (by about $10$ units) and remain close 
to each other (about $-53 \pm 2$ units),
which is close to the mean values computed with the more 
complete set of $8$ species (about $-52 \pm 1$ units).

Therefore, the mean specific heat at constant pressure for 
sea salts may be set to 
\begin{align}
\overline{C}_{p{\rm s}} & \approx\, 
 -\,52 \pm 1~\mbox{J~K${}^{-1}$~mol${}^{-1}$}
 \label{eq_Cp_seasalt_molar} \:  
\end{align}
or, with $M_{\rm s}$ given in (\ref{Eq_Ms}),  
\begin{align}
\overline{C}_{p{\rm s}} & \approx\, 
 -\,1656 \pm 30~\mbox{J~K${}^{-1}$~kg${}^{-1}$}
 \label{eq_Cp_seasalt_kg} \: . 
\end{align}
The specific entropy of sea salts at $0{}^{\circ}$C 
can then be computed from the value of 
$\eta_{\rm s0}$ at $25{}^{\circ}$C given by 
(\ref{Eq_eta_s_25C_kg}), leading to 
\begin{align}
\!\!
\eta_{\rm s0}\mbox{($0{}^{\circ}$C)} 
& \approx ( 1488.3 \pm 13 ) 
\nonumber \\
& \quad - ( 1656 \pm 30 )
\times \ln\!\left(\frac{273.15}{298.15}\right) ,
\label{Eq_eta_s_0C_Marquet_2023_T} \\
\!\!
\eta_{\rm s0}\mbox{($0{}^{\circ}$C)} & \approx
1633.3 \pm 15~\mbox{J~K${}^{\,-1}$~kg${}^{\,-1}$} \; .
\!\!
\label{Eq_eta_s_0C_Marquet_2023}
\end{align}

 \subsection{The TEOS10 absolute seawater entropy}
\label{subsection_abs_entropy_TEOS10}
\vspace*{-2mm}

Therefore, it is possible to compute the absolute seawater 
entropy by considering the absolute values for the 
reference entropies $\eta_{\rm s0}$ and $\eta_{\rm w0}$, 
to be taken into account in the standard TEOS10 
value $\eta\:(x,y,z) = \eta^{\rm W}_{\rm Fei03}\:(y,z) 
+ \eta^{\rm S}_{\rm Fei08}\:(x,y,z)$ recalled in 
(\ref{eq_TEOS10_eta_x_y_z}), 
with the pure-water and salinity parts 
$\eta^{\rm W}_{\rm Fei03}\:(y,z)$ and 
$\eta^{\rm S}_{\rm Fei08}\:(x,y,z)$ shown in 
Tables~\ref{Table_TEOS10_sigma_W_x_y_z} 
and \ref{Table_TEOS10_sigma_S_x_y_z}, respectively. 

Both values of the constant terms $\eta_{\rm w0}$ and 
$-g_{010}/40 \approx 0.147\,64$ in 
$\eta^{\rm W}_{\rm Fei03}\:(y,z)$ have no physical impact
(zero differential, and thus zero time derivatives, 
gradients, and turbulent fluxes).
Differently, the impact of the first-order salinity 
term in $\eta^{\rm S}_{\rm Fei08}\:(x,y,z)$, recalled in the 
first line of Table~\ref{Table_TEOS10_sigma_S_x_y_z}, 
can be summarised by the increment term
\begin{align}
\Delta \eta_s
& = (\eta_{\rm s0}-\eta_{\rm w0}) \times 
\frac{(S_{\rm A}-S_{\rm SO})}{1000} 
\label{Eq_etas_minus_etaw} \;
\end{align}
(in~J~K${}^{\,-1}$~kg${}^{\,-1}$) 
and have physical impacts on time derivative, 
gradients, and turbulent fluxes of the seawater 
entropy.
This increment must be understood as a way to compute 
the absolute entropy from the standard TEOS10 
value, according to
\begin{align}
\eta_{\rm abs} & = \eta_{\rm std/TEOS10}  \;+\; \Delta \eta_{\rm s}  
\label{Eq_Delta_eta_ans_std} \; .
\end{align}
The underlying assumption is first to trust the standard TEOS10 formulation 
to provide the entropy variations to be valid around the standard 
conditions, where $\eta_{\rm std/TEOS10}$ arbitrarily cancels out.  
The next step is to amend this standard formulation by adding the 
absolute-entropy increment $\Delta \eta_{\rm s}$ with respect to 
$S_{\rm SO}$, which also cancels out for these standard conditions 
without loss of generality.

The present values for $\eta_{\rm s0}$ and $\eta_{\rm w0}$ 
due to the arbitrary hypothesis 
$\eta(S_{\rm SO},t_{\rm SO},p_{\rm SO})=0$ 
made in TEOS10 can be replaced 
by the third-law absolute values at $0{}^{\circ}$C 
given in 
(\ref{Eq_eta_s_0C_Marquet_2023}) and 
(\ref{eq_eta_w0_NEA_TDB_1992}): 
\begin{align}
\eta_{\rm s0} - \eta_{\rm w0} 
& \approx (1633.3 \pm 15) - (3513.4 \pm 1.7) \; 
\nonumber 
\end{align}
and
\begin{align}
\Delta \eta_{\rm s}(S_{\rm A})
& \approx (-1880 \pm 17)
\times \frac{(S_{\rm A}-S_{\rm SO})}{1000} 
\label{Eq_etas_minus_etaw_value} \; 
\end{align}
expressed in J~K${}^{\,-1}$~kg${}^{\,-1}$, 
with $S_{\rm SO}=35.165\,04$~g~kg${}^{-1}$ 
and with an uncertainty in $\Delta \eta_{\rm s}$ of only $1$~\%.


The absolute-entropy increment (\ref{Eq_etas_minus_etaw_value}) 
is large in comparison with the present (arbitrary) entropy 
value recalled in the first line of $\eta^{\rm S}_{\rm Fei08}\:(x,y,z)$ 
shown in Table~\ref{Table_TEOS10_sigma_S_x_y_z},
with $x^2=S_{\rm A}/40.188\,617$ and 
$g_{210} \approx 168.072\,408\,311\,545$
leading to 
$-\,g_{210} \times (x^2/40) \approx
 -104.6 \times (S_{\rm A} / 1000)$.  
This is about $6$\% of the value of 
$\Delta \eta_{\rm s}$ given by  
(\ref{Eq_etas_minus_etaw_value}) and only 
$6$ times the uncertainty in $\Delta \eta_{\rm s}$.

\begin{figure*}[htb]
 \includegraphics[width=0.45\linewidth,angle=0,clip=true]{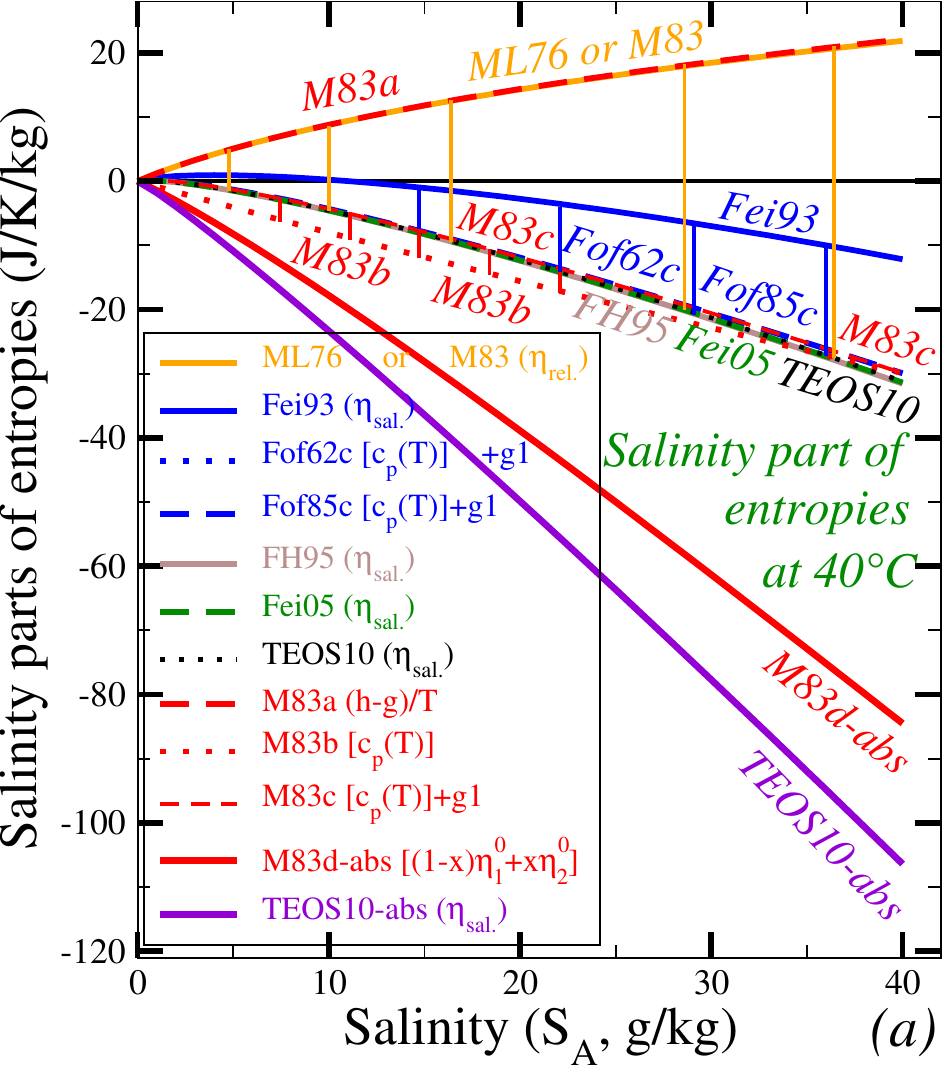}
 \includegraphics[width=0.45\linewidth,angle=0,clip=true]{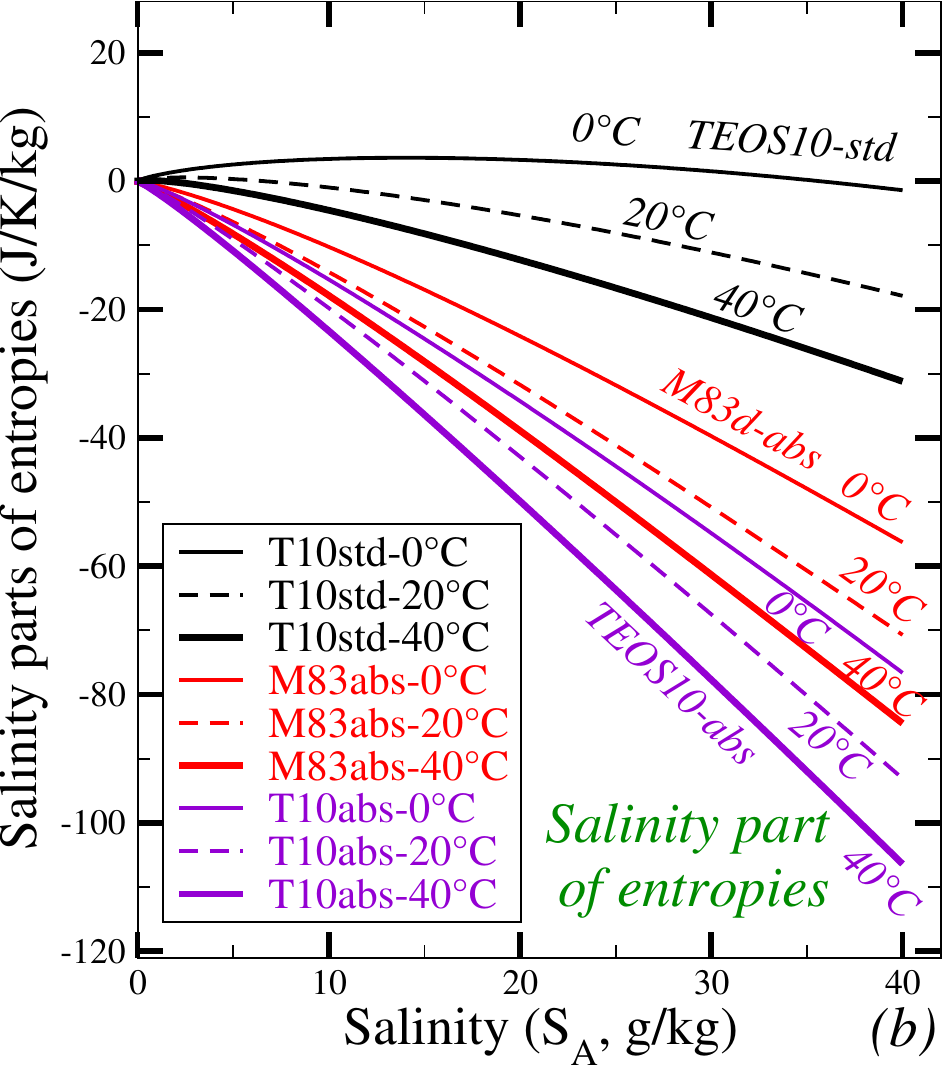} \\
 \includegraphics[width=0.45\linewidth,angle=0,clip=true]{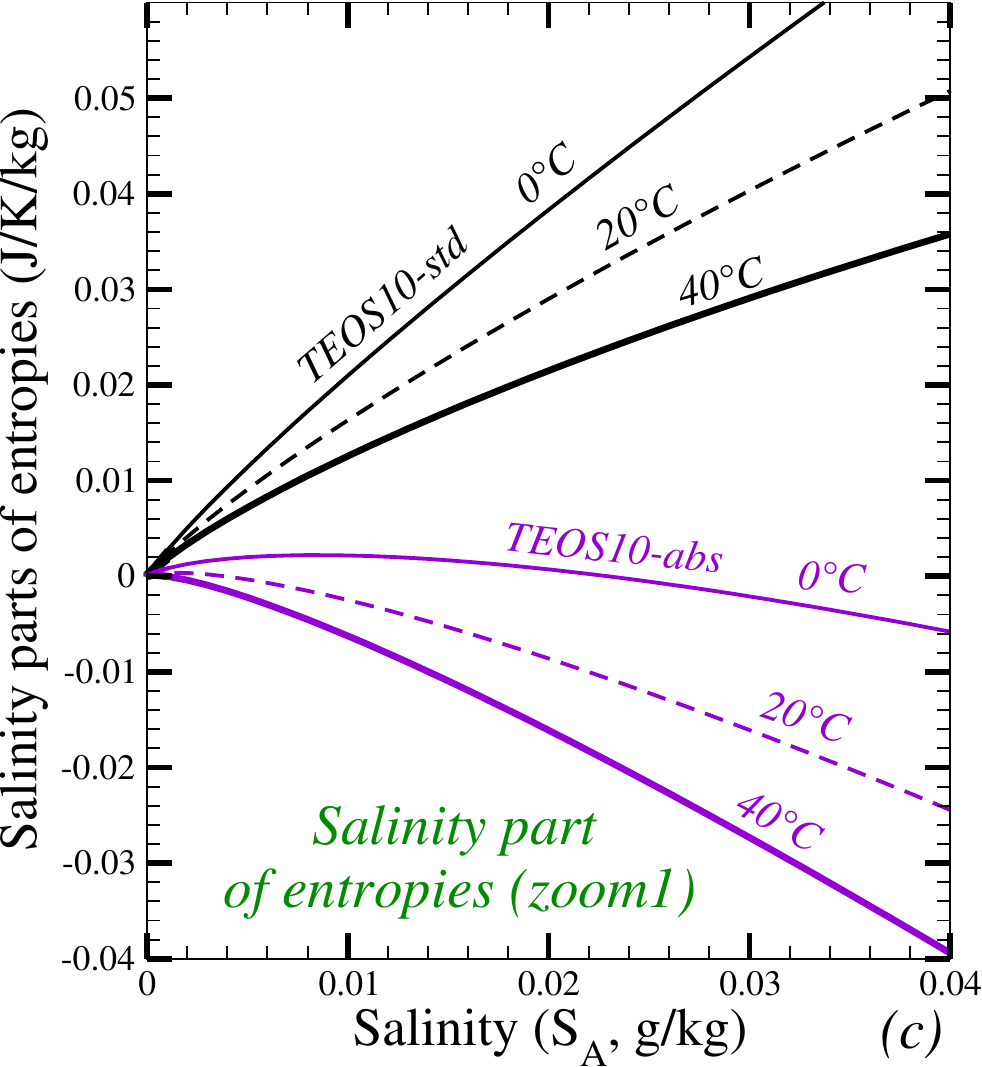}
 \includegraphics[width=0.45\linewidth,angle=0,clip=true]{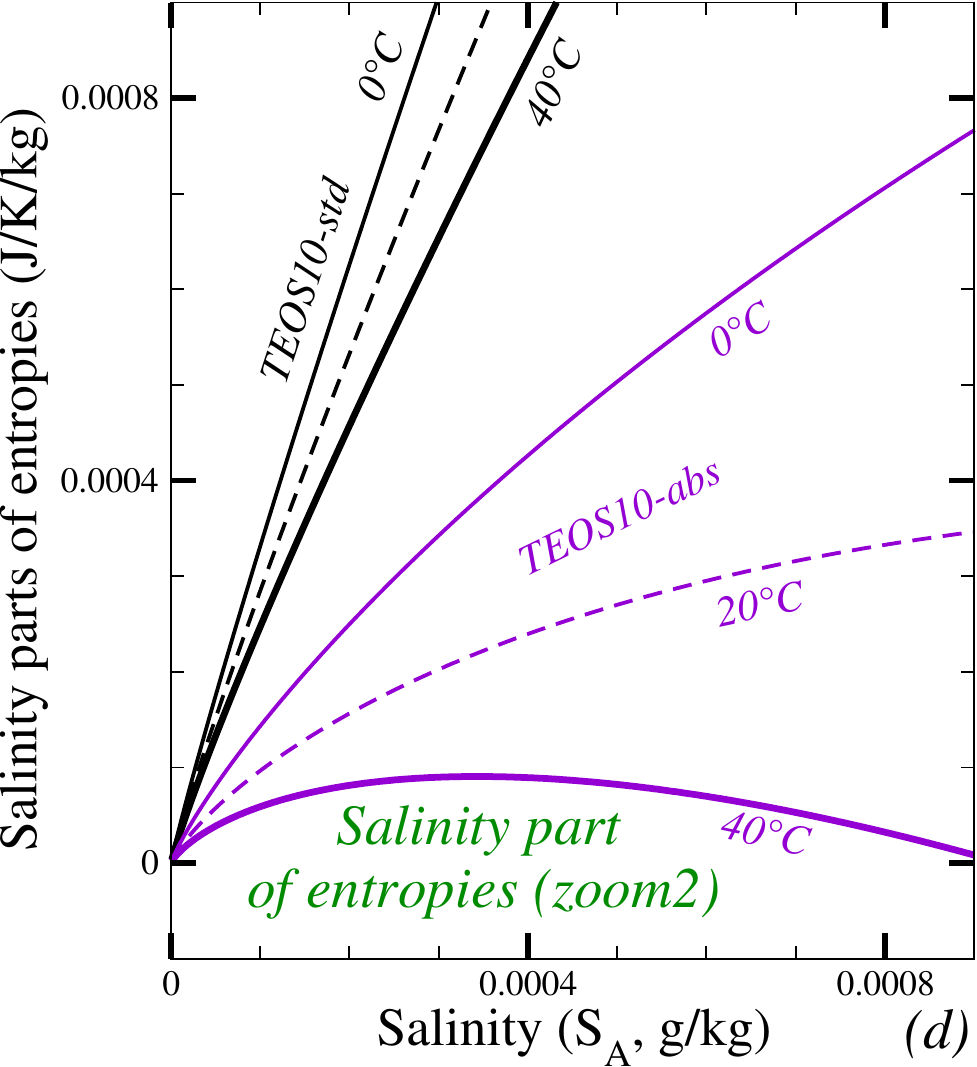} \\
\vspace*{-6mm}
\caption{{\it\small 
In (a):
the `relative' or salinity parts of the
seawater entropies plotted against the absolute salinity 
at the temperature $40$~${}^{\circ}$C 
and for several old and more recent formulations
 \citep[][Fof62]{Fofonoff_1962}, 
 ML76+M83, M83a, M83b, M83c, M83d-abs 
 (from ML76 and ML83, see the main text) 
 \citep[][Fof85]{Fofonoff_JGR_1985}, 
 \citep[][Fei93]{Feistel_1993}, 
 \citep[][FH95]{Feistel_Hagen_1995}, 
 \citep[][Fei05]{Feistel_Ocean_Sci_2005}, 
 \citep[][TEOS10]{Feistel_TEOS_manual_2010}, 
 \citep[][TEOS10-abs with $\Delta \eta_{\rm s}$,  
 Eq.~(\ref{Eq_etas_minus_etaw_value})]{Feistel_TEOS_manual_2010}.
In (b):  
the salinity parts of the seawater entropies plotted against  
the absolute salinity for three selected temperatures  
($0$, $20$ and $40$~${}^{\circ}$C) and for TEOS10, M83d-abs and TEOS10-abs. 
In (c) and (d):  
two zoomed versions of (b) for TEOS10 and TEOS10-abs and for very 
small salinity, to show the impact for small $x^2=S_{\rm A}/40$ 
of the term $x^2\:\ln(x)$ corresponding to 
$(S_{\rm A}/40)\:\ln(S_{\rm A}/40)\,/\,2$. 
\label{Fig_Sanility_parts_Seawater_Entropies}}\\
-----------------------------------------------------------------------------------------------------------------------------------------------
}
\end{figure*}
\clearpage 

 \section{Numerical applications}
\label{section_numerical_applications}
\vspace*{-2mm}

 \subsection{Comparisons with the Millero and others entropies}
\label{subsection_abs_entropy_Millero}
\vspace*{-2mm}

It became apparent during the review process that it was necessary to describe in some detail the comparisons between the more recent Feistel's and TEOS10 formulations and the older papers of ML76 and M83,
where for the first time an attempt was published to take into account the absolute values of the entropies of pure water and ocean salts to calculate, in some ways, the absolute seawater entropy.
In particular, it was necessary to explain how entropy is calculated in Millero's papers in a somewhat atypical way, in a so-called ``{\it\,relative\,}'' way, which explains the significant differences compared to more recent formulations.

Figs.~\ref{Fig_Sanility_parts_Seawater_Entropies} show the salinity parts 
at constant pressure departure ($p=0$) of the seawater entropies 
$\eta_{\rm sw/sal.}(t,S_{\rm A}) = \eta_{\rm sw}(t,S_{\rm A}) - \eta_{\rm w}(t)$  
defined as the increment from the liquid-water part $\eta_{\rm w}(t)$. 
For the specific seawater entropy written like in (\ref{eq_TEOS10_eta_x_y_z}), 
the salinity part $\eta_{\rm sw/sal.}(t,S_{\rm A})$ can be computed as follows: 
\begin{align}
\!\!\!\!\!
\eta_{\rm sw}(t,S_{\rm A}) 
& = \left(1\:-\:\frac{S_{\rm A}}{1000}\right) \: \eta_{\rm w}(t)
\nonumber \\
& \; + \left(\frac{S_{\rm A}}{1000}\right) \: \eta_{\rm s}(t,S_{\rm A})
\label{Eq_eta_salinity_1} \: , \\
\!\!\!\!\!
\eta_{\rm sw}(t,S_{\rm A}) 
& = \eta_{\rm w}(t)
\nonumber \\
& \; + \left(\frac{S_{\rm A}}{1000}\right)  
\left[\:\eta_{\rm w}(t)-\eta_{\rm s}(t,S_{\rm A})\:\right]
\label{Eq_eta_salinity_2} , \!\! \\
\!\!\!\!\!
\eta_{\rm sw/sal.}(t,S_{\rm A}) & 
= \left(\frac{S_{\rm A}}{1000}\right)  
\left[\:\eta_{\rm s}(t,S_{\rm A})-\eta_{\rm w}(t)\:\right]
\label{Eq_eta_salinity_3} . \!\!\!\!
\end{align}
Consequently, the impact of reference entropy values on $\eta_{\rm sw/sal.}$ 
is the quantity $(S_{\rm A}/1000)\:(\eta_{\rm s0}-\eta_{\rm w0})$, which  
is the part computed in (\ref{Eq_etas_minus_etaw}) 
and (\ref{Eq_etas_minus_etaw_value}) independent of 
the TEOS10 value $S_{\rm SO} =35.165\,04$~g~kg${}^{-1}$.

\begin{table*}[hbt] 
\caption{\it\small The low-order Gibbs' coefficients 
$g_0=g[1;0;0]$, $g_1=g[1;1;0]$, $g[2;0;0]$ 
and $g[2;1;0]$ involved in Feistel's relationships  
in the terms $(g_0+g_1\:y)\:x^2\:\ln(x)\:$, 
$g[2;0;0]\:x^2$ and $g[2;1;0]\:x^2\:y$ for
the non-dimensional variables 
$x^2=S_P/(40$\:g\:kg${}^{\,-1})$ or 
$x^2=S_A/(40.188\,617$\:g\:kg${}^{\,-1})$
and $y=t/\mbox{($40{}^{\circ}$C)}$.}
\vspace*{2mm}
\centering
\begin{tabular}{cccccccccc}
\hline 
 & $g[2;0;0]$
 & $g[2;1;0]$
 & $g[1;0;0]$
 & $g[1;1;0]$
\\ \hline  
 \citet[][Fei93]{Feistel_1993}
& [\,$-4204.5207$\,]
& [\,$-615.7087$\,]
&    $5814.9808$ 
&    $851.5440$
\\   
 \citet[][FH95]{Feistel_Hagen_1995}
& ($1531.9856$)
& ($160.9033$)
&  $5813.3468$ 
&  $851.3047$
\\  
 \citet[][Fei05]{IAPWS_Feistel_2003,Feistel_Ocean_Sci_2005}
& ($1376.280...$) 
& ($140.577...$)
&  $5813.287...$
&  $851.2959...$
\\ 
\citet[][TEOS10]{Feistel_TEOS_manual_2010} 
& ($1416.276...$) 
& ($168.072...$)
&  $5812.815...$
&  $851.2267...$
\\ \hline 
\end{tabular}
\label{Table_gijk_coef}
\end{table*}

The three Feistel's and TEOS10 curves labeled FH95, Fei05, and TEOS10 are clearly 
very close to each other in the Fig.~\ref{Fig_Sanility_parts_Seawater_Entropies}(a).
The older formulation of \citet[][in blue, labelled Fei93]{Feistel_1993} differs 
from these three almost superimposed curves, with a positive bias (vertical blue thin lines) 
but with similar negative and decreasing values for $S_{\rm A} > 10$~g~kg${}^{-1}$.

This positive bias can be explained by the different arbitrary definitions made for the zero of enthalpies and entropies in both Fei93 and \citet[][hereafter FH94]{Feistel_Hagen_Thermo_Ocean_1994}.
This was recalled in FH95, where it was explicitly stated (p.~269) that the arbitrary choice made in Fei93 (and still retained in FH94) was to  ``{\it\,define the enthalpy per salt particle to vanish at infinite dilution\,}'' (i.e., $S=0$~PSU), and to arbitrarily decide from FH95 onwards ``{\it\,to set entropy and enthalpy to zero for the standard ocean state $t=0{}^{\circ}$C, $p=0$~MPa and $S=35$~PSU\,}'' because ``{\it\,entropy per salt particle diverges when salinity goes to zero\,}'' in Fei93 and FH94.

I show in Table~\ref{Table_gijk_coef} a comparison of the lower-order terms 
$(g[1;0;0]+g[1;1;0]\:y)\:x^2\:\ln(x) + (g[2;0;0]+g[2;1;0]\:y)\:x^2$ 
for the Gibbs function $g_{\rm sw}$, which enter the entropy formula via 
$\eta_{\rm sw} = -\,\partial g_{\rm sw}/\partial t = 
 -(1/40)\,\partial g_{\rm sw}/\partial y$, and thus via 
$-(1/40)\:[\:\,g[1;1;0]\:x^2\:\ln(x) + g[2;1;0]\:x^2\:]$. 
If the varying coefficient $g[2;0;0]$ only concerns the enthalpy, 
the other varying coefficient $g[2;1;0]$ impacts the value of the seawater 
entropy, and Fei93's value is indeed very different from those for 
FH95, Fei05, and TEOS10, with a large impact on the seawater entropy 
via the term $-\:[\:g[2;1;0]/1.6\:]\:(S_{\rm A}/1000)$. 
This means from (\ref{Eq_eta_salinity_3}) that the changes in 
$-\:g[2;1;0]/1.6$ correspond to those in $\eta_{\rm s0}-\eta_{\rm w0}$ 
computed at $0{}^{\circ}$C. 
The difference of about $160.9+615.7 = 776.6$ between Fei93's and FH95's   
formulations corresponds to a change in $\eta_{\rm s0}-\eta_{\rm w0}$ of about 
$-485.7$~J~K${}^{-1}$~kg${}^{-1}$, and from (\ref{Eq_eta_salinity_3}) 
to a change in $\eta_{\rm sw/sal.}(t,S_{\rm A})$ of about 
$-19.4$~J~K${}^{-1}$~kg${}^{-1}$ at $S_{\rm A}=40$~g~kg${}^{-1}$, 
which is in agreement with Fig.~\ref{Fig_Sanility_parts_Seawater_Entropies}(a) 
and with the same vertical shift of about $-19.4$~J~K${}^{-1}$~kg${}^{-1}$ 
(from $-12.164$ to $-31.561$) between Fei93's and FH95's curves. 

The Millero curves (ML76 or M83) plotted in orange on Fig.~\ref{Fig_Sanility_parts_Seawater_Entropies}(a) 
show an even greater vertical shift, with values made positive and increasing for all values of $S_{\rm A}$.
We could try to interpret this increased shift relative to the curves of Fei93, FH95, Fei05, and 
TEOS10 as a result of the absolute definitions of reference entropies a priori included in ML76 and M83. 
However, the impact of these absolute definitions is negative when using the formula 
(\ref{Eq_etas_minus_etaw_value}) established above for $\Delta \eta_{\rm s}$, as shown by the 
differences between the black curve for TEOS10 (standard) and the purple curve for TEOS10-abs. 
It therefore becomes important to be able to explain the reasons for such unusual features for 
the impacts in ML76 and M83 of the absolute definitions of reference entropies.

It can first be noted that the orange and increasing curve  
for ML76 or M83 has already been plotted by 
\citet[][Eq.~44, p.~373, Fig.~16, p.~374]{Sharqawy_Lienhard_Zubair_2010}
and \citet[][Fig.~5.1, p.~125]{Qasem_Thermo_Prop_Saline_Water_2023},     
who already showed that all other `specific salinity' 
curves plotted from  
\citet{Chou_Thesis_Thermo_aqueous_1968},
\citet{Pitzer_Peiper_Busey_1984},
\citet[][IAPWS]{IAPWS_Sup_Sat_Press_St_Petersburg_1992}, 
\citet{Sun_al_Deep_Sea_Res_2008},  
\citet[][IAPWS]{Cooper_Dooley_IAPWS_2008}, and 
\citet{Nayar_al_2016} were decreasing with 
the salinity $S_{\rm A}$.  
This means that the increasing orange curve for ML76+M83 
in Figure~\ref{Fig_Sanility_parts_Seawater_Entropies}(a) 
is truly atypical, as it differs from all the others 
(Chou-1968, Pitzer-1984, IAPWS-1992, 
Feistel-1993-2003-2005,
Sun-2008, IAPWS-2008, TEOS-2010, Nayar-2016). 
Conversely, the decreasing black curve for TEOS10 (standard), 
which I relied on in my paper, is very similar to all 
the others except those by Millero.

In fact, it can be shown that the formulations proposed by 
ML76 (Eq.~136, p.~1072) and M83 (Eq.~147, p.~36)
corresponding to the curves in orange (ML76 or M83) do not 
include the absolute values of the reference entropies.  
Indeed, they do not correspond to the definitions (\ref{Eq_eta_salinity_1}) 
and (\ref{Eq_eta_salinity_2}) leading to the saline parts 
(\ref{Eq_eta_salinity_3}) for the entropy of seawater. 
In order to prove this, it can be noted that the plot of the curve 
M83a (in bold dashed red) for the quantity $(h-g)/T$ is superimposed 
on that of the entropy $\eta$ (ML76+M83), with $h$ the enthalpy and 
$g=h-T\:\eta$ the Gibbs function previously defined both in 
ML76 (Eq.~67, p.~1050 and Eq.~118, p.~1068) and 
M83 (Eqs.~54-57, p.~15-16 and Eqs.~126-129, p.~32). 
These definitions were derived long before the absolute liquid-water and sea-salt 
entropies $\overline{\eta}_1^{\,\circ}$ and $\overline{\eta}_2^{\,\circ}$ 
were respectively defined in ML76 (p.~1070-1072) and M83 
(p.~34-35)$\,$\footnote{$\:$Note that I have not used the notations $s$, $S$, 
$\overline{S}^{\,\circ}$, $\overline{S}_1^{\,\circ}$, 
$\overline{S}_2^{\,\circ}$, ..., used in Millero's papers for the 
entropies, with the same letter $S(\mbox{\textperthousand})$ 
used to denote the salinity and with an obvious 
risk of confusion in ML76 and M83.}. 
This confirms that neither ML76, M83, or M83a can represent 
the absolute version of the saline part of seawater entropy.

In order to continue the investigation, it is possible to start from the definition 
of the specific heat of seawater expressed in ML76 (Eqs.~52, p.~1044) 
and M83 (Eqs.~34-37, p.~11) 
as 
\begin{align}
\!\!
c_{p{\rm sw}}(t,S) & = c_{p0}(t) 
  + A_{{\rm c}p}(t) \: S + B_{{\rm c}p}(t) \: S^{3/2}
  \label{Eq_M83_Cpsw} \: . \!\! 
\end{align}
\begin{align}
\!\!\!\!\!\!
A_{{\rm c}p}(t)  & = -7.644 + 0.10727 \: t 
                     -1.38\,10^{-3} \; t^2  
  \label{Eq_M83_Acp} \: , \\
\!\!\!\!\!\!
B_{{\rm c}p}(t)  & =  0.177 - 4.08\,10^{-3} \: t 
                   + 5.35\,10^{-5} \; t^2  
  \label{Eq_M83_Bcp} \: ,  \\
\!\!\!\!\!\!
c_{p0}(t) & =  4217.4 - 3.720283\: t 
            + 0.1412855 \; t^2
  \nonumber \\
\!\!\!\!\!\!
  & \;   + 2.654387\,10^{-3} \; t^3 
         + 2.093236\,10^{-5} \; t^4
  \label{Eq_M83_cp0} \: . \!\!
\end{align}
The next step is to define a value of the seawater entropy 
by integrating $c_{p{\rm sw}}(t)/(t+273.15)$ up to an arbitrary function 
of salinity $\eta_0$ only, and independent of the temperature. 
This leads, with $\eta_0\mbox{($0{}^{\circ}$C)} = 0$, to the red dotted curve M83b,  
with a small negative bias with respect to the FH95, Fei05, and TEOS10 
curves (thin vertical red lines). 

This small negative bias of the curve M83b is clearly reduced with the red curve M83c plotted with the following alternative choice for $\eta_0\mbox{($0{}^{\circ}$C)}$, obtained by adding the impact of the Gibbs ideal-solution term $g_1\:x^2\:\ln(x)\:y$ with the value $g_1=g[1;1;0] \approx 851.3$ typical of FH95+Fei05+TEOS10, for which the impact on entropy is $-[\:(851.3/1600)\:\ln(S/40)\:]\:S >0$ for $0<S<40$, and null for $S=0$ and $S=40$.
The same method can be used by integrating the values of $c_{p{\rm sw}}(t)/(t+273.15)$ published in the earlier articles of \citet[][Eqs.~29-30-31, p.~13-14)]{Fofonoff_1962} and \citet[][Table-7, p.~3339]{Fofonoff_JGR_1985}, with the blue curves Fof62c and Fof85c obtained by adding the same term $-(g_1/40)\:x^2\:\ln(x)$ with $g_1 \approx 851.3$ as for M83c. 

As a result, all curves for Fof62c, M83c, Fof85c, FH95, Fei05, and TEOS10 
almost overlap in Fig.~\ref{Fig_Sanility_parts_Seawater_Entropies}(a).  
This means that the ML76+M83 formulation of $c_{p{\rm sw}}(t,S)$ recalled 
in (\ref{Eq_M83_Cpsw}) is relevant, and that the atypical behavior 
of the M83 and M83b curves must correspond to a deeper difference in 
the functions $h$, $g=h-T\:\eta$ and $\eta$ defined in these papers.
Furthermore, \citet[][p.~102-103]{Feistel_1993} indicates that 
several problems exist in the previous Millero's formulations 
(forming parts of what he called the `EOS80' system of equations): 
``{\it\, ... there is a numerical mismatch in the paper of ML76, 
in transforming heat capacity to apparent molal heat capacity 
in the form of a generalized Debye-H\"uckel formula. 
Because this formula is used later to compute enthalpy and 
free enthalpy} (and thus entropy), 
{\it the temperature dependence of both is inconsistent with 
heat capacity. 
... theory predicts the power series in salinity would begin 
with $S\:\sqrt{S}$ in the case of relative enthalpy and with 
$S\:\ln(S)$ in the case of relative free enthalpy} 
(and thus of entropy). 
{\it The corresponding functions reported are smoothed 
out at small salinities to begin simply with $S$. 
Therefore they cannot be used correctly to compute, say, 
the chemical potential} (and thus the entropy).
{\it ... while density etc. are defined in absolute terms, 
thermal properties are only given relative to an infinitely 
diluted reference state. 
It is not obvious how these quantities have to be combined 
correctly if necessary, for example when computing the 
chemical potential} (and thus the entropy). 
{\it `Relative' means that $G(S,t,p)$ is zero at the infinite dilution 
reference state, where all ion-ion interactions are neglected at 
temperature $t=0{}^{\circ}$C ($T= 273.15$~K) and pressure $p=0$~bar 
($P=1$~atm).\,}''

Indeed, it turns out that the ML76 and M83 definitions of 
`relative' and `relative apparent' parts of any extensive 
quantity $S$ (like the heat content or enthalpy, the volume, 
the available enthalpy or Gibbs function, and the entropy 
noted $S$) 
are the same as those published in the previous paper 
by \citet[][]{Millero_Hansen_Hoff_h_seawater_JMR_1973},  
and in fact go back to the older books by LR61 and 
\citet[][]{Harned_Owen_3rd_1958,Harned_Owen_2nd_1950,Harned_Owen_1st_1943}.
The `relative' feature means considering only the deviations 
$S_{\rm rel.} = S-S^{\circ}$ from the molar or specific actual value 
$S  = n_1\:S_1 + n_2\:S_2$ and the `infinite-solution' value 
$S^{\circ} = n_1\:S^{\circ}_1 + n_2\:S^{\circ}_2$. 
They are both computed for the actual $n_1$ moles of liquid water 
and $n_2$ moles of sea salts, and with the `infinite-solution' 
values $S^{\circ}_1$ and $S^{\circ}_2$ corresponding to 
$n_2 \rightarrow 0$ and thus $S \rightarrow 0$. 
It is thus clear that the omission of a function $S^{\circ}$ for entropy--which 
varies with salinity--renders this 'relative' view of entropy, defined by 
$S(t,S_{\rm A},p)-S^{\circ}(t,S_{\rm A},p)$, somewhat meaningless 
for the study of the entropy function $S(t,S_{\rm A},p)$.
This is why \citet[][Eq.5-9, p.~96]{Chou_Thesis_Thermo_aqueous_1968} 
defined the specific entropy of a solution consisting of 
$n_1$ mole of water and $n_2$ moles of salt as 
$s = n_1 \:\overline{s}^{\circ}_1 + n_2\:\overline{s}^{\circ}_2 
+ (...)$,  
therefore including the term 
$n_1 \:\overline{s}^{\circ}_1 + n_2\:\overline{s}^{\circ}_2 $
not included in Millero's `relative' formulations. 
Similarly, \citet[][Eq.1, p.~3]{Pitzer_Peiper_Busey_1984} 
defined the Gibbs function of the real system as 
$G = n_1 \:G^{\circ}_1 + n_2\:\overline{G}^{\circ}_2 
+ (...)$, again including the terms 
not included in `relative' Millero's formulations.

Starting from (\ref{Eq_eta_salinity_1}) the `infinite-solution' 
and `relative' value of the seawater entropy (noted $\eta$) 
can be written as 
\begin{align}
\!\!\!\!
\eta_{\rm sw}^{\circ}(t,S_{\rm A}) 
& = \left(1\:-\:\frac{S_{\rm A}}{1000}\right) \: \times \: \eta_{\rm w}(t)
\nonumber \\ 
\!\!\!\!
& \quad + \left(\frac{S_{\rm A}}{1000}\right) \: \times \: \eta_{\rm s}(t,S_{\rm A}=0)
\label{Eq_eta_infinite_sol} \: , \\ 
\!\!\!\!
\eta_{\rm sw/rel.}(t,S_{\rm A}) & 
  =  \eta_{\rm sw}(t,S_{\rm A}) 
\:-\:\eta_{\rm sw}^{\circ}(t,S_{\rm A}) 
\label{Eq_eta_relat_1} \: , \!\!\\ 
\!\!\!\!
\eta_{\rm sw/rel.}(t,S_{\rm A}) & 
= \eta_{\rm w}(t) 
+ \left(\frac{S_{\rm A}}{1000}\right) \: \times \: 
\nonumber \\ 
\!\!\!\!
& \quad  
 \left[\: \eta_{\rm s}(t,S_{\rm A})
    \:-\: \eta_{\rm s}(t,S_{\rm A}=0)
 \:\right]
\label{Eq_eta_relat_2} . \!\!
\end{align}
It should be emphasised that the infinite-solution entropy 
$\eta_{\rm sw}^{\circ}(t,S_{\rm A})$ in (\ref{Eq_eta_infinite_sol}) 
is a variable quantity depending on salinity through the concentrations 
$1-S_{\rm A}/1000$  and $S_{\rm A}/1000$ of the real solution, 
even though  
$\eta_{\rm w}^{\circ}(t)=\eta_{\rm w}(t)$ and 
$\eta_{\rm s}^{\circ}(t)=\eta_{\rm s}(t,S_{\rm A}=0)$ 
calculated in infinite-dilution mode do not depend on the salinity 
and only on the temperature.  

It therefore appears that the so-called `relative' entropy of ML76 and M83 is 
very different from the salinity parts studied in Fei93, FH95, Fei05, and TEOS10, 
with the liquid-water entropy $\eta_{\rm w}(t)$ in the last term of 
the salinity-part formula (\ref{Eq_eta_salinity_3}) replaced by 
the infinite-solution sea-salts entropy $\eta_{\rm s}(t,S_{\rm A}=0)$ 
in the `relative-entropy' formula (\ref{Eq_eta_relat_2}), 
together with the additional first term $\eta_{\rm w}(t)$. 

According to ML76 (p.~1072) and M83 (p.~36),  
``{\it\,Since future workers may wish to look at entropy surfaces 
in the oceans,\,}'' the `relative-entropy' formula was fitted  
to the function $[\:h(t,S)-g(t,S)\:]/(t+273.15)$ to give  
\begin{align}
\!\!\!\!  
\eta_{\rm sw/rel.}(t,S) & = A_s(t) \; S(\mbox{\textperthousand})  
     \;+\; B_s(t) \; S(\mbox{\textperthousand})^{3/2} 
\nonumber \\ 
\!\!\!\! & \quad  
     \;+\; C_s(t) \; S(\mbox{\textperthousand})^{2}  
\label{Eq_M83_s} \: , \\
A_s(t) & = 
  \;+\; 1.42185 
  \;-\; 3.1137\,10^{-4} \; t
\nonumber \\ 
\!\!\!\! & \quad 
  \;+\; 4.2446\,10^{-6} \; t^2
\label{Eq_M83_s_A} \: , \\
B_s(t) & = 
  \;-\; 0.21762 
  \;+\; 4.1426\,10^{-4} \; t
\nonumber \\ 
\!\!\!\! & \quad 
  \;-\; 1.6285\,10^{-6} \; t^2
\label{Eq_M83_s_B} \: , \\
C_s(t) & = 
  \;+\; 1.0201\,10^{-2} 
  \;+\; 1.5903\,10^{-5} \; t
\nonumber \\ 
\!\!\!\! & \quad 
  \;-\; 2.3525\,10^{-7} \; t^2
\label{Eq_M83_s_C} \: . 
\end{align}
However, this formula cannot be used to compute and plot a salinity part of the seawater entropy 
as in Fig.~\ref{Fig_Sanility_parts_Seawater_Entropies}(a).
To show this, it is possible to determine the function of temperature and salinity that must be added to this `relative' value of ML76 and M83 to calculate the saline part of seawater entropy.
Indeed, the use of (\ref{Eq_eta_infinite_sol})-(\ref{Eq_eta_relat_2}) allows the computation of 
\begin{align}
\eta_{\rm sw/sal.}(t,S_{\rm A}) & 
\; = \;  \eta_{\rm sw}(t,S_{\rm A}) 
\:-\:\eta_{\rm w} (t)
\label{Eq_eta_sal_2} \: , \\ 
\eta_{\rm sw/sal.}(t,S_{\rm A}) & 
\; = \; \eta_{\rm sw/rel.}(t,S_{\rm A})
 \:+\: 
 \left(\frac{S_{\rm A}}{1000}\right) 
 \; \times 
\nonumber \\
& \; \quad 
 \left[\: \eta_{\rm s}(t,S_{\rm A}=0) 
        - \eta_{\rm w} (t)
 \:\right]
\label{Eq_eta_sal_3} \: . 
\end{align}
The last term of (\ref{Eq_eta_sal_3}), which is absent from $\eta_{\rm sw/rel.}(t,S_{\rm A})$, should be able to explain the fact that the ML76 or M83 curves (in orange) and the M83a curve (in dashed red) are so different from the others. 
They are in particular the only ones to be increasing with salinity.
The formulation (\ref{Eq_eta_sal_3}) confirms that absolute values are not taken into account in the adjustments (\ref{Eq_M83_s})-(\ref{Eq_M83_s_C}), since they are absent from the two `relative' functions $h(t,S)$ and $g(t,S)$ previously defined and on which these adjustments are based. 
However, it is possible to partly calculate these two missing terms,  
$\eta_{\rm s}(t,S_{\rm A}=0)$ and $\eta_{\rm w} (t)$, 
where the reference values 
$\eta_{\rm s0}(25\mbox{${}^{\circ}$C})$ and $\eta_{\rm w0}(25\mbox{${}^{\circ}$C})$ 
are involved, respectively. 

To do so, the ML76+M83 specific absolute reference sea-salt 
entropy is first recalled in (\ref{Eq_ML76_M83_eta_s0_kg}). 
Then, the entropy of liquid water $\eta_{\rm w} (t)$ was computed by 
ML76 (p.~1072) and M83 (p.~36) by first fitting the specific 
heat capacity $c_{pw}(t)$ given by (\ref{Eq_M83_cp0}) 
in terms of the absolute temperature $T~\mbox{(K)}$. 
This leads to 
\begin{align}
c_{pw}(T) & = 
        21.02343\,10^{3} 
  \;-\; 119.8696 \; T 
\nonumber \\ & \quad 
  \;+\; 9.828769\,10^{-2} \; T^2  
\nonumber \\ & \quad 
  \;+\; 8.466352\,10^{-4} \; T^3 
\nonumber \\ & \quad 
  \;-\; 1.554175\,10^{-6} \; T^4 
\label{Eq_M83_Cpw} \: ,
\end{align}
with an accurate specific heat capacity of about   
$4217.0$~J~K${}^{\,-1}$~kg${}^{\,-1}$ at $273.15$~K 
in (\ref{Eq_M83_Cpw}), 
versus $4217.4$~J~K${}^{\,-1}$~kg${}^{\,-1}$ at $0{}^{\circ}$C
in (\ref{Eq_M83_cp0}).
Then, by using the integral  
\begin{align}
\!\!\!
\eta_w(T) \;=\; \eta_w(\mbox{$25{}^{\circ}$C}) 
\;+\; \int_{298.15}^{T} \frac{c_{pw}(T')}{T'}\: dT'
\label{Eq_M83_S_w_int} \: , \!\!
\end{align}
ML76 and M83 obtained the liquid-water entropy formula:  
\begin{align}
\eta_{w}(T) & = \: \left[\: 
        \eta_{\rm w0}(\mbox{$25{}^{\circ}$C}) 
  \;-\; 92.8218 \: \right] 
\nonumber \\ & \quad 
  \;+\; 21.02343\,10^{3} \: \ln(T) 
  \;-\; 119.8696 \; T  
\nonumber \\ & \quad 
  \;+\; 4.914384\,10^{-2} \; T^2 
  \;+\; 2.822117\,10^{-4} \; T^3 
\nonumber \\ & \quad 
  \;-\; 3.885437\,10^{-7} \; T^4 
\label{Eq_M83_Sw} \: .
\end{align}
The bracketed terms were erroneously set to $+22.8218$ 
in ML76 and M83, which would correspond to the wrong value  
$\eta_{\rm w0}(\mbox{$25{}^{\circ}$C}) 
\approx 115.64$~J~K${}^{\,-1}$~mol${}^{\,-1}$. 
This value is very different from the absolute reference entropy 
of about $70.00$~J~K${}^{\,-1}$~mol${}^{\,-1}$ recalled 
in (\ref{Eq_ML76_M83_eta_w0_mol}), published in LR61 (p.~137)
and duly retained in ML76 and M83. 
The first mistake in ML76 and M83 is therefore the sign 
of the sum of bracketed terms to be set to $-22.8218$ 
to possibly agree with the molar entropy 
$70.00$~J~K${}^{\,-1}$~mol${}^{\,-1}$. 
However, the second mistake is to use this molar value 
in the relationship (\ref{Eq_M83_Sw}) that is valid for 
a specific value expressed in J~K${}^{\,-1}$~kg${}^{\,-1}$, 
with therefore the need to use instead the specific value 
$\eta_{\rm w0}(\mbox{$25{}^{\circ}$C}) 
\approx 3885.65$~J~K${}^{\,-1}$~kg${}^{\,-1}$ 
recalled in (\ref{Eq_ML76_M83_eta_w0_kg}) 
and computed with the molar mass 
$M_1 \approx 18.015$~g~mol${}^{\,-1}$ for \ch{H2O}
recalled p.~1074 in ML76.  
This leads to a bracketed term of about 
$+3792.83$~J~K${}^{\,-1}$~kg${}^{\,-1}$ 
in (\ref{Eq_M83_Sw}) and provides a relevant 
formula for the specific liquid water entropy 
$\eta_{w}(T)$ to be included in the 
last term of (\ref{Eq_eta_sal_3}).

The next problem is that the entropy of sea salts $\eta_{\rm s}(t,S_{\rm A})$,  
depending on the temperature and salinity, is not calculated explicitly by ML76 or M83, 
like that $\eta_{\rm w}(t)$ of liquid water in (\ref{Eq_M83_Sw}) and that relative 
$\eta_{\rm sw/rel.}(t,S)$ for seawater in (\ref{Eq_M83_s}).
However, the absolute value of the reference sea-salt entropy at infinite dilution 
$\eta_{\rm s0}\mbox{($25{}^{\circ}$C)} \approx 1513.1$~J~K${}^{-1}$~kg${}^{-1}$ 
recalled in (\ref{Eq_ML76_M83_eta_s0_kg}) was explicitly calculated by ML76 and M83 
at $25{}^{\circ}$C.
We can therefore imagine a first-order formulation to be used in (\ref{Eq_eta_sal_3}) 
and based on (\ref{Eq_eta_s_0C_Marquet_2023_T}), but valid for variable absolute 
temperatures $T$ and based on the specific heat 
$\overline{C}_{p{\rm s}}\mbox{($25{}^{\circ}$C)}  
 \approx -\,1656$~J~K${}^{-1}$~kg${}^{-1}$ 
calculated in (\ref{eq_Cp_seasalt_kg}), leading to 
\begin{align}
\!\!\!\!
\eta_{\rm s}(t,S_{\rm A}=0) &  
\; \approx \; \eta_{\rm s0}\mbox{($25{}^{\circ}$C)} 
\nonumber \\ & \quad 
+ \overline{C}_{p{\rm s}}\mbox{($25{}^{\circ}$C)}
\times \ln\!\left(\frac{T}{298.15}\right) 
\label{Eq_eta_s_0C_Marquet_2023_T_bis} 
\: . \!\!
\end{align}
By substituting (\ref{Eq_M83_Sw}) and (\ref{Eq_eta_s_0C_Marquet_2023_T_bis}) into
(\ref{Eq_eta_sal_3}), we obtain in Fig.~\ref{Fig_Sanility_parts_Seawater_Entropies}(a) 
the M83d-abs curve in solid red that is decreasing with salinity. 
This new absolute entropy M83d-abs curve is relatively close to the TEOS10-abs curve (in solid purple) 
and is very different from the ML76, M83, and M83a (increasing) curves for the `relative' entropies. 

This M83d-abs curve corresponds to the expected and hoped-for result, 
as it validates, in a way, the absolute definition of seawater entropy 
as I propose and have plotted with TEOS10-abs. 
The M83d-abs curve corrects the `relative' ML76+M83+M83a aspect 
that was too atypical and was not included in the subsequent IAPWS 
and TEOS10 projects, in which Millero has nevertheless participated 
since 1983. 
The difference between the M83d-abs and TEOS10-abs curves can probably 
be explained by the remaining approximations and imperfections in the 
ML76 and M83 formulations, as listed by \citet[][p.~102-103]{Feistel_1993}
about the `EOS80' system of equations.

It should be noted that the decrease in entropy with salinity is 
more pronounced for absolute versions and 
is not contrary to the fact that entropy must increase during the process 
of dissolving salts in pure water. 
As an example, the process of dissolution of crystals of NaCl 
corresponds to a positive change in seawater entropy for the reaction 
\ch{NaCl}${}_{\rm \,cr}$\ch{->}(Na${}^{+}$)${}_{\rm aq}$+(Cl${}^{-}$)${}_{\rm aq}$.
Indeed, the entropy for a crystal of \ch{NaCl} is 
$72.14$~J~K${}^{\,-1}$~mol${}^{\,-1}$, whereas 
the entropy for the separate ions (Na${}^{+}$)${}_{\rm aq}$ 
and (Cl${}^{-}$)${}_{\rm aq}$ is
$59.00 + 56.36 = 115.36$~J~K${}^{\,-1}$~mol${}^{\,-1}$. 
This leads to an increase of entropy of 
$115.36-72.14=+43.22$~J~K${}^{\,-1}$~mol${}^{\,-1}$,  
representing the impact on the entropy of the 
dissolution of the \ch{NaCl} crystal. 
In fact, the decrease in absolute entropy with salinity must be interpreted 
differently: it is as a decrease in the average entropy per unit mass when a 
mass of water with a higher salinity is added to another with a lower salinity, 
with the salts already dissolved in both masses of water. 
For such a process the entropy decreases because the reference entropy of ocean 
salts ($1513.1$~J~K${}^{-1}$~kg${}^{-1}$) is smaller than that of liquid water 
($3885.65$~J~K${}^{\,-1}$~kg${}^{\,-1}$).
In this sense, the errors in the tuning leading to the ML76 and M83 `relative'  
formulation had to be corrected in the present formulation based on the 
amended version of TEOS10, with, moreover, the use of the absolute values 
of the reference entropies.

The second plot in Fig.~\ref{Fig_Sanility_parts_Seawater_Entropies}(b) 
shows the joint variations with the salinity of the curves 
$\eta_{\rm sw}(t,S_{\rm A})$ for a selection of three temperatures: 
$t=0{}^{\circ}$C, $20{}^{\circ}$C, and $40{}^{\circ}$C.
We can see that the behaviors observed at $40{}^{\circ}$C remain unchanged at 
other temperatures, with a more noticeable decrease in all cases 
in Figs.~\ref{Fig_Sanility_parts_Seawater_Entropies}(b)(c)(d) 
for the absolute versions of the entropies.

Finally, the impact of the term $x^2\:\ln(x)$ in the TEOS10's entropy 
formulation corresponding to the term $S_{\rm A}\:\ln(S_{\rm A})$ 
introduced by \citet[][Eq.~1.1.2, p.~2689]{Onsager_Fuoss_1932}  
is to create a vertical tangent at the origin. This is particularly
visible in Figs.~\ref{Fig_Sanility_parts_Seawater_Entropies}(c) and (d),    
where I show that this behaviour starts to exist only for 
$S_{\rm A}$ much smaller than $1$~g~kg${}^{\,-1}$. 
Differently, a general decrease of the absolute entropy exists 
for the whole oceanic range of salinity from 
$5$ to $40$~g~kg${}^{\,-1}$.

 \subsection{The temperature-salinity $(t-S_{\rm A})$ diagram}
\label{subsection_t_S_diagram}
\vspace*{-2mm}

\begin{figure*}[htb]
\centering
\includegraphics[width=0.55\linewidth,angle=0,clip=true]{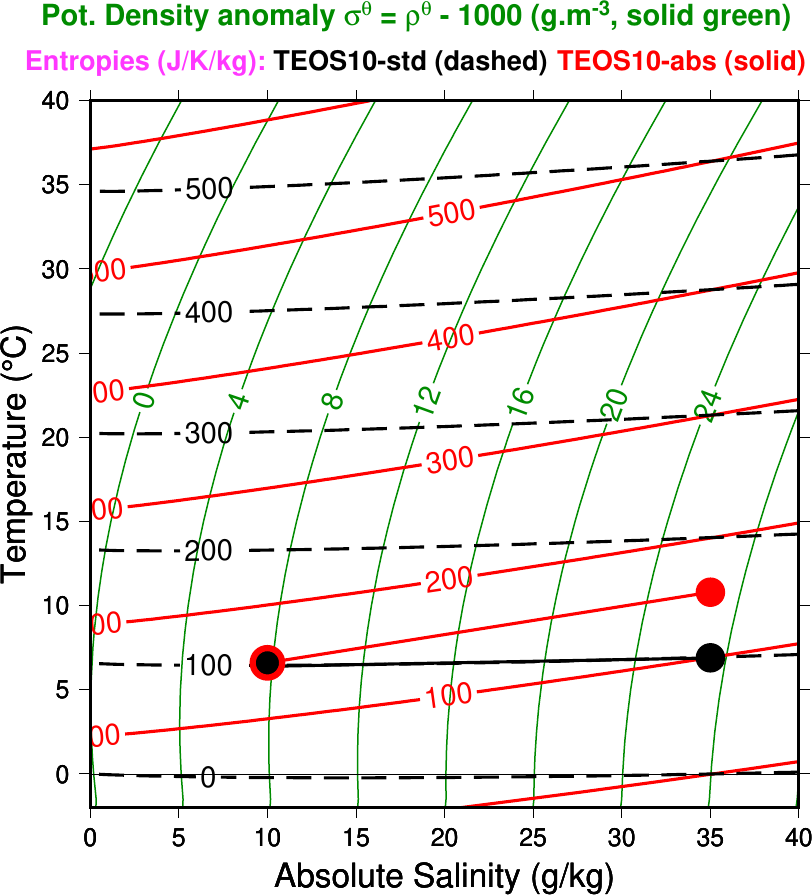} \\
\vspace*{-3mm}
\caption{{\it\small 
The isopycnic potential-density anomaly (solid green) lines, 
the standard TEOS10 entropy (dashed black) lines, and 
the new absolute TEOS10 entropy (solid red) lines 
are plotted on a classical $t-S_{\rm A}$ Temperature-Salinity diagram.
The three red and black circles represent isentropic processes  
(see the main text). 
\label{Fig_temperature_salinity_diagram}}\\
-----------------------------------------------------------------------------------------------------------------------------------------------
}
\end{figure*}

The same results are obtained with the study of the 
other temperature-salinity $(t-S_{\rm A})$ diagram plotted in the 
 Figs.~\ref{Fig_temperature_salinity_diagram}.       
The green lines are for constant values of the TEOS10 
potential-density  
anomaly$\,$\footnote{$\:$Note that $\sigma^{\theta}$ 
is computed with the TEOS10's function  
\texttt{gsw\_pot\_rho\_t\_exact(sa,t,p,pr=0)}
with the pressure changed to a fixed reference 
sea pressure anomaly $p_r=0$ in an isentropic and isohaline 
manner \citep[see][p.~28]{Feistel_TEOS_manual_2010}.}
($\sigma^{\theta}=\rho^{\theta} - 1000$, in g~m${}^{-3}$).
Using potential density instead of local density will make 
it easier to compare the density of points of vertical profile 
at different depths in the second part of the paper, always 
using the same family of green lines (surface, $p_r=0$).  

In this $t-S_{\rm A}$ diagram, the new absolute seawater entropy 
red lines are different from the standard TEOS10 black lines: 
the new absolute values increase with the salinity, 
whereas the standard TEOS10 values are almost constant
(especially for the smaller values of salinity and 
temperature).
Similar differences (not shown) exist between the standard and 
absolute versions of the seawater entropy plotted 
with the 
entropy-temperature diagram published in 
\citet[][Fig.4, p.~103]{Feistel_2010}.

Three black and red circles are plotted to show the differences 
in temperature associated with an isentropic process starting at 
about $6.6{}^{\circ}$C and with a change in salinity from $10$ to 
$35$~g~kg${}^{\,-1}$: the change in temperature is about $4{}^{\circ}$C 
greater for the absolute version than for the standard TEOS10 
version ($\Delta t \approx +0.3{}^{\circ}$C compared with $+4.2{}^{\circ}$C). 
This difference is significant, as shown by the applications 
to real cases described in Part~II.

It should be noted here (like in Part II) that the term ``\,isentropic\,'' 
(or same entropy) includes possible joint variations in pressure, 
temperature, and salinity, without prejudging any adiabatic 
(no exchange of heat) or closed (same salinity) aspect of 
the evolution of ocean parcels. 
This is analogous to the definition of ``\,isopycnal\,'' processes, 
where density (or potential density) remains constant regardless of 
joint variations in pressure, temperature, and salinity, with variations 
in temperature and salinity along isopycnals described with spiciness.

 \subsection{General impacts of the Third law in physics}
\label{subsection_General_impacts}
\vspace*{-2mm}

It may be important to recall the first validations 
for \ch{H2}, \ch{Ar} and \ch{Hg} made close to $0$~K and  
recalled by \citet[][p.~185-186]{Nernst_new_heat_theorem_1926}  
for the Sackur-Tetrode-Planck absolute entropy constants 
    defined and studied by 
\citet[][]{Planck1917,Nernst_Theor_Exp_Neuen_Warmesatzes_1918,
Nernst_1921} and 
\citet[][]{Planck_Vorlesungen_Warmestrahlung_4th_1921_ge}.  
These authors showed that the theoretical and statistical-quantum 
absolute value $C_0 \approx -1.61$ was in agreement with the 
experimental and calorimetric third-law value 
$C_0 \approx -1.62 \pm 0.03$. 
This justified and allowed the computation of the saturation 
pressures from the third-law absolute entropies if the latent 
heats are known (see section~12.1 in the SM). 

\begin{figure*}[htb] 
\vspace*{1mm}
\includegraphics[width=0.49\linewidth,angle=0,clip=true]{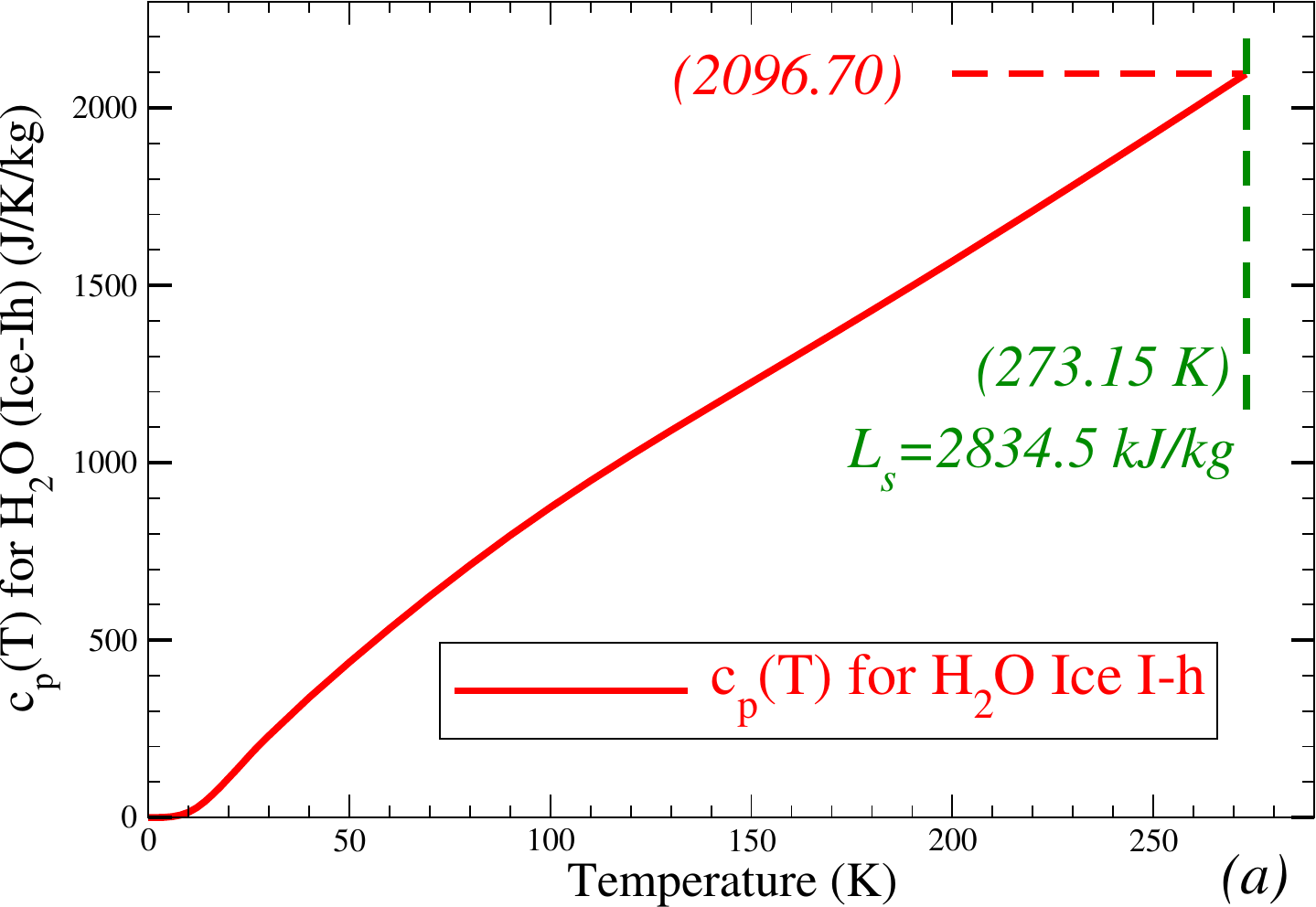}
\vspace*{1mm}
\includegraphics[width=0.49\linewidth,angle=0,clip=true]{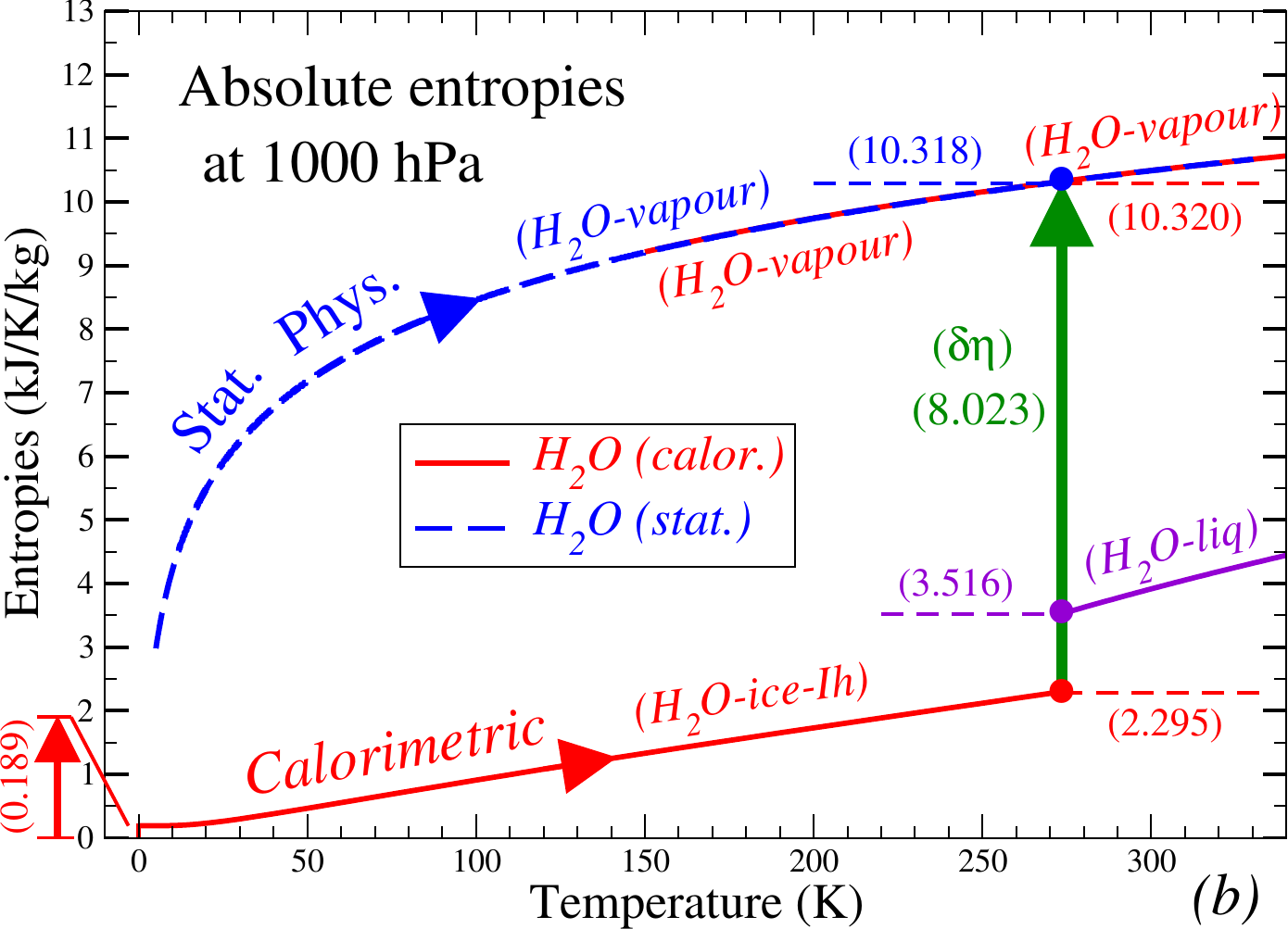}
\vspace*{-3mm}
\caption{{\it\small 
(a): The specific heat at constant pressure $c_p(T)$ for \ch{H2O} (Ice-Ih) for absolute temperatures $T$ from $0$~K to $T_0=273.15$~K. 
(b): The absolute entropies for \ch{H2O} (Ice-Ih, liquid and vapour) from $0$~K to $340$~K, with the calorimetric values $\eta_0 +\int_0^T c_p(T')\:d\ln(T')$ (solid red lines) including the Pauling-Nagle residual entropy $\eta_0 \approx 0.189$~kJ~K\,${}^{-1}$~kg\,${}^{-1}$ at $0$~K, and the statistical values $\eta=k\:\ln(W)$ (dashed blue line) automatically taking into account the residual entropy at $0$~K and the latent heat of sublimation $L_{\rm sub}(T_0) \approx 2834.5$~kJ~kg\,${}^{-1}$ at $273.15$~K. 
The term $\delta \eta \approx 10.318-2.295 = 8.023$~kJ~K\,${}^{-1}$~kg\,${}^{-1}$ (green arrow) is the difference between the statistical water-vapour and  calorimetric Ice-Ih absolute entropies at $T_0=273.15$~K.
\label{Fig_Entropy_H2O_stat_calor}}\\
-----------------------------------------------------------------------------------------------------------------------------------------------
}
\end{figure*}

It is possible to show another modern validation of 
the theoretical and calorimetric third-law absolute values 
computed at $0{}^{\circ}$C for the water-vapour and ice-Ih entropies.
I have first computed (see section~4.6 and Eq.~126 in the SM) the 
third-law statistical-quantum water-vapour entropy 
\begin{align}
\!\!\!\!
\eta_{\,\rm v{/\rm 3rd}}^{\rm\,stat.}(T_0,p_0) 
\;\approx\; 10317.9 \pm 0.63\mbox{~J~K\,${}^{-1}$~kg\,${}^{-1}$}    
\label{Eq_Entropy_H20_vap_stat}  \!\!
\end{align}
via the relationship $\eta=k\:\ln(W)$ represented by the 
blue arrow and the path up to the blue disk in  
Fig.~\ref{Fig_Entropy_H2O_stat_calor}(b). 
Here, $W$ corresponds to the translational, rotational,  
and vibrational degrees of freedom of the 3D molecule 
\ch{H2O}. 
I have also computed, in a completely independent way 
(see section~3.5 and p.~19 in the SM), the absolute 
calorimetric entropy for \ch{H2O} ice-Ih at $0{}^{\circ}$C
\begin{align}
\eta_{\,\rm i{/\rm 3rd}}^{\rm\,calor.}(T_0) 
\;\approx\; 2295.5 \pm 1.9\mbox{~J~K\,${}^{-1}$~kg\,${}^{-1}$}    
\label{Eq_Entropy_H20_ice_calor} \; . 
\end{align}
I have used the relationship $\eta_0 +\int_0^T c_p(T')\:d\ln(T')$ 
represented by the red arrow and the path up to 
the red disk in Fig.~\ref{Fig_Entropy_H2O_stat_calor}(b), 
with the experimental values of $c_p(T)$ plotted 
in Fig.~\ref{Fig_Entropy_H2O_stat_calor}(a)
and where $\eta_0 \approx 189$~J~K\,${}^{-1}$~kg\,${}^{-1}$ 
is the Pauling-Nagle residual entropy at $0{}^{\circ}$CK. 

As a consequence, a new validation of the third law of thermodynamics 
can be obtained by computing 
(see section~3.5 and p.~19 in the SM)
the calorimetric water-vapour 
entropy at $T_0=273.15$~K and $p_0=10^5$~Pa via the relationship
\begin{align}
\eta_{\,\rm v/{\rm 3rd}}^{\rm\,calor.}(T_0,p_0) 
& \; = \;  
\eta_{\,\rm i{/\rm 3rd}}^{\rm\,calor.}
\left[\:T_0,\,p_{\rm sat}(T_0)\:\right] 
\nonumber \\ & \quad 
+ \frac{L_{\rm sub}(T_0)}{T_0} 
- R_{\rm v}\,\ln\!\left[\:\frac{p_0}{p_{\rm sat}(T_0)}\:\right] 
\nonumber \\ & \quad 
\approx 
10320 \pm 4\mbox{~J~K\,${}^{-1}$~kg\,${}^{-1}$}
\label{Eq_Entropy_H20_vap_calor} \; , 
\end{align}
where $L_{\rm sub}(T_0) \approx 2834.5 \pm 0.5$~kJ~kg\,${}^{-1}$
is the latent heat of sublimation, 
$R_{\rm v} \approx 461.52 \pm 0.02$~J~K\,${}^{-1}$~kg\,${}^{-1}$ 
the gas constant for water vapour and 
$p_{\rm sat}(T_0) \approx 611.15 \pm 0.10$~Pa  
the water-vapour saturation pressure over ice-Ih 
at $0{}^{\circ}$C. 
Clearly, this water-vapour absolute third-law calorimetric 
(\ref{Eq_Entropy_H20_vap_calor}) and the previous 
statistical-quantum entropies (\ref{Eq_Entropy_H20_vap_stat}) 
are in close agreement, up to $0.02$~\%. 

Conversely, one can use (\ref{Eq_Entropy_H20_vap_stat}) 
and (\ref{Eq_Entropy_H20_ice_calor}) to calculate either 
the latent heat of sublimation $L_{\rm sub}(T_0)$ or the saturation pressure 
$p_{\rm sat}(T_0)$ variable from the other variable and from the 
difference in third-law entropies  
\begin{align}
\delta \eta\,(T_0,p_0) 
& \:=\: 
  \eta_{\,\rm v/{\rm 3rd}}^{\rm\,stat.}(T_0,p_0) 
\:-\: 
  \eta_{\,\rm i/{\rm 3rd}}^{\rm\,calor.}
  \left[\:T_0,\,p_{\rm sat}(T_0)\:\right] 
\nonumber \\ 
& \;\approx\; 
8022.4 \pm 2.6\mbox{~J~K\,${}^{-1}$~kg\,${}^{-1}$}
\label{Eq_delta_Entropy_H20}
\end{align}
(i.e., the green arrow in Fig.~\ref{Fig_Entropy_H2O_stat_calor}(b) 
in between the red and blue disks) by using the relationships 
\begin{align}
\!\!\!
& L_{\rm sub}(\delta \eta,\,p_{\rm sat},\,T_0,\,p_0) 
\;=\;  
\nonumber \\ 
\!\!\!
& \quad \quad T_0 \:
\left\{\: \delta \eta\,(T_0,p_0) \:+\: 
 R_{\rm v}\:\ln\!\left[\:\frac{p_0}{p_{\rm sat}(T_0)}\:\right] 
 \:\right\}  
\label{Eq_H20_Ls_T0}  
\end{align}
or
\begin{align}
\!\!\!
& p_{\rm sat}(\delta \eta,\,L_{\rm sub} ,\,T_0,\,p_0)
\;=\; 
\nonumber \\ 
\!\!\!
& \quad \quad p_0 \: 
 \exp\!\left[\: \frac{\delta \eta\,(T_0,p_0)}{R_{\rm v}} 
 \:-\:  \frac{L_{\rm sub} (T_0)}{R_{\rm v} \:T_0} \:\right] 
\label{Eq_H20_Psat_T0}  
\end{align}
to arrive at $L_{\rm sub}(T_0) \approx 2833.9 \pm 0.8$~kJ~kg\,${}^{-1}$ 
or $p_{\rm sat}(T_0) \approx 608.4 \pm 6$~Pa, which are indeed 
in agreement with the experimental values
$L_{\rm sub}(T_0) \approx 2834.5 \pm 0.5$~kJ~kg\,${}^{-1}$ and 
$p_{\rm sat}(T_0) \approx 611.15 \pm 0.10$~Pa.  

These calculations show the predictive power of the statistical 
(for water vapour) and calorimetric (for Ice-Ih) third-law independent 
computations leading to the blue and red disks, respectively. 
These calculations form a validation of the physical significance 
of the third-law values of the entropy by forming a little-known  
link between the numerical values of 
$L_{\rm sub}(T_0)$ and $p_{\rm sat}(T_0)$, which are generally considered 
to be independent experimental constants.
This creates an unexpected third-law implicit impact on the 
thermodynamic conditions at the surface of the oceans and 
thus impacts the measurable evaporation processes.

Incidentally, these calculations may also provide a justification for 
and a physical interpretation of the residual entropy of 
$189$~J~K\,${}^{-1}$~kg\,${}^{-1}$.
Indeed, if we subtract this value from the absolute entropy of ice, 
then according to (\ref{Eq_delta_Entropy_H20}), the quantity 
$\delta \eta\,(T_0,p_0)$ increases from about $8023$ 
to about $8212$~J~K$^{-1}$~kg$^{-1}$, 
resulting in, according to (\ref{Eq_H20_Ls_T0}) and (\ref{Eq_H20_Psat_T0}), 
the unrealistically high values of 
$L_{\rm sub} (T_0) \approx 2885.7$~kJ~kg$^{-1}$ and 
$p_{\rm sat}(T_0) \approx 917.6$~Pa 
in comparison with the experimental values of 
$L_{\rm sub}(T_0) \approx 2834.5 \pm 0.5$~kJ~kg\,${}^{-1}$ and 
$p_{\rm sat}(T_0) \approx 611.15 \pm 0.10$~Pa.
The bias of $189 \times 273.15 \approx 51.6$~kJ~kg$^{-1}$ for $L_{\rm sub} (T_0)$ 
and the multiplicative factor of $\exp(189/461.65) \approx 1.506$ for 
$p_{\rm sat}(T_0)$ are indeed unrealistic values.
This means that it was not possible to separate the residual value from the 
absolute calorific entropy of ice-Ih, as suggested by \citet{Feistel_2019}  
who tried ``{\it Distinguishing between Clausius, Boltzmann and
Pauling entropies of frozen non-equilibrium states.\,}''
This also means that this quantity, $189$~J~K$^{-1}$~kg${}^{-1}$, 
is likely hidden somewhere in the quantum-statistical calculations 
carried out by Sackur, Tetrode, Planck, and others for the 
translational, rotational, and vibrational degrees of freedom 
of atoms and molecules.

Moreover, I recall in section~12.2 of the SM that the equilibrium constants 
of chemical reactions $K(T)$ ultimately depend on the absolute entropies 
$S^0$ of the reactants and products, via the well-known stability relationship 
   $\Delta G_{\rm r} = \Delta H_{\rm r} - T \:\Delta S^0 = -R\:T\,\ln(K)$. 
General forms like 
   $\ln(K) = \Delta S^0/R - \Delta H_{\rm r}/(R\:T) = A + B/T + C\:\ln(T)$ 
are considered both in ozone  
chemistry \citep[NASA-JPL publication by][]{Burkholder_al_O3_Nasa_2020} and 
seawater chemistry, where it was calculated in \citet{Weiss_CO2_1974} 
by adjustments to the observed data. 
They are also considered, for instance, by \citet{Millero_CO2_1995} 
for the study of carbon dioxide concentration in the oceans and more generally 
described in the \citet[Departement of Energy,][]{DOE_Dickson_Goyet_1994} 
publication. 
In the general form above, the third-law absolute entropies of reactants 
and products must impact the constant term $A$. 
I show in section~12.2 of the SM such an impact for 
the atmospheric reactions \ch{Cl + O2 <-> ClO2} and \ch{NO + NO2 <-> N2O3} 
and the seawater reaction \ch{HCO3- <-> H+ + {CO3}$\!\!{}^{2\,-}$} 
(ionization of the bicarbonate anion). 
The constant terms $A$ 
can be computed from the theoretical or experimental values of the 
third-law absolute entropies of molecules, anions, and cations acting 
as reactants and products.
Therefore, the concentrations of ozone in the atmosphere and of sea salts 
in the oceans both depend on the third-law absolute reference entropies 
as those considered in the present paper, in Millero's papers, and in 
some few others in atmospheric science
\citep[like][...]{Hauf_Hoeller_1987,Marquet2011QJ}, via the photochemistry 
of the stratosphere and the electrolytic chemistry of the oceans.
This is how the absolute values of the entropies influence the observed 
values of the concentrations of ozone and sea salts, where the 
concentrations vary with the temperature and pressure. 
This influences, in return and via the interactions with the radiations, 
the vertical profiles of the temperature, which is an obvious observable 
quantity.

Moreover, I show in section~12.9 of the SM that similar impacts exist 
for the turbulence acting on the observable temperature-like variables. 
This, according to \citet[][]{Richardson_19a,Richardson_22}, should be 
calculated via a turbulent mixing acting on the absolute entropy variable 
and not on the temperature variable (see the section 
``{\it A need to update the seawater entropy\:}''). 
I show in section~12.9 of the SM that the Lewis number (ratio of the thermal 
and water exchange coefficients) is different from unity for the 
moist-air absolute entropy variable $\theta_{\rm s}$ \citep{Marquet2011QJ}, 
for which $K_{\rm s} > 0$.  
This generates a counter-gradient term for the usually used liquid-water 
potential temperature $\theta_{\rm l}$ variable \citep{Betts1973}, 
for which $K_{\rm h}$ does not even have a definite sign. 
This means that the vertical and horizontal structures of temperature 
must ultimately depend on the absolute definition of the difference 
in entropy (like $s_{\rm vr}-s_{\rm dr}$ for the atmosphere and 
$\eta_{\rm s0} - \eta_{\rm w0}$ in the ocean) and the corresponding 
absolute-entropy exchange coefficients. 
Accordingly, it is possible to make a link between these atmospheric 
turbulent phenomena and those in the oceans. 
Indeed, several of the parameterizations are based on the principle of 
''{\it double diffusion,\,}'' first studied by \citet{Stern_Salt_Fountain_1960} 
for the ''{\it Salt-Fountain\,}'' and still considered in \citet{Ma_Peltier_2024} 
for the polar oceans. Here, the exchange coefficients are different for 
the large-scale velocity (momentum), heat (temperature), and salinity, 
respectively 
\citep[see, for instance,][]{Canuto_Howard_Cheng_Dubuvikov_Ocean_Turb_II_2002}. 

Moreover, in order to comply with Richardson's requirements, 
an equivalent of the atmospheric variable 
    $\theta_{\rm s} = T_0 \: \exp[\:(s-s_{\rm d0}) /c_{p{\rm d}}\:]$ 
\citep{Marquet2011QJ,Marquet_Stevens_JAS22} 
to be used in turbulent schemes of oceanic simulations 
could be defined as the seawater absolute-entropy (potential?) temperature
$\theta_\eta$ derived from the value of the absolute entropy $\eta_{\rm abs}$ 
given by (\ref{Eq_etas_minus_etaw}), (\ref{Eq_Delta_eta_ans_std}),   
and (\ref{Eq_etas_minus_etaw_value}), but rewritten as 
   $\eta_{\rm abs} = c_{\rm w} \: \ln(\theta_\eta/T_0)$,  
and thus with 
   $\theta_\eta = T_0 \: \exp(\eta_{\rm abs}/c_{\rm w})$ 
written as  
\begin{align}
\!\!\!\!
\theta_\eta & = 273.15 
  \times \exp\left(\frac{
            \eta_{\rm std/TEOS10}}{4218}
         \right) 
\nonumber \\
\!\!\!\!
& \quad \times 
  \exp\left[
     (-0.446 \pm 0.004) 
     \left(\frac{S_{\rm A}-S_{\rm SO}}{1000}\right)
  \right] 
,  \label{eq_theta_eta_new} \!\!
\end{align}
where $T_0=273.15$~K, $c_{\rm w} 
\approx 4218$~J~K${}^{\,-1}$~kg${}^{\,-1}$ 
and $S_{\rm SO}=35.165\,04$~g~kg${}^{-1}$.
Note that the additive true constant term $\eta_{\rm w0}$ is 
discarded in the definition (\ref{eq_theta_eta_new}) 
that should be
   $\theta_\eta = T_0 \: \exp[\:(\eta_{\rm abs}-\eta_{\rm w0})/c_{\rm w}\:]$ 
instead, simply because $\eta_{\rm w0}$ has already been discarded 
in the TEOS10 definition of the pure-water part $\eta^{\rm W}_{\rm Fei03}$ 
recalled in Table~\ref{Table_TEOS10_sigma_W_x_y_z} 
(and only there in the present paper, as a true constant).
This definition (\ref{eq_theta_eta_new}) for 
$\theta_\eta$ leads to exactly the same properties 
as the entropy $\eta_{\rm abs}$ because 
both $c_{\rm w}$ and $T_0$ are constant.
However, $\theta_\eta$ (like $\theta_{\rm s}$ for the atmosphere) 
is different from the associated entropy variable in that 
it has the same dimension (Kelvin) as the absolute 
temperature ($T$), the potential temperature ($\theta$),  
and the conservative temperature ($\Theta$) already provided
as outputs of the TEOS10 software.  

I show in Figs.~42 to 44 in section~12.9 of the SM that 
$\theta$ and $\Theta$ remain very close to each other for several 
arctic and tropical CTD profiles (up to $\pm 0.03{}^{\circ}$C). 
The differences between the actual temperature $T$ and 
both $\theta$ and $\Theta$ are also small close to the surface 
(up to $\pm 0.03{}^{\circ}$C for depths smaller than $250$~dbar), where  
the impact of salinity on $\theta_\eta$ is the largest 
and can reach $\pm 1.5{}^{\circ}$C. 
Note the interesting feature that $\theta_\eta$ becomes 
similar to both $\theta$ and $\Theta$ for layers deeper 
than $3000$~dbar (up to $\pm 0.04{}^{\circ}$C), where the differences 
with $T$ can reach $-0.6{}^{\circ}$C. 
These numerical evaluations confirm the potential of the 
$\theta_\eta$ variable to become a kind of thermal variable 
on which ocean turbulence is acting and becoming a kind of 
''{\it absolute seawater potential temperature\,}''. 
This variable $\theta_\eta$ is numerically similar 
to $\theta$ and $\Theta$ in deeper layers, 
but different from them in the mid and surface layers. 
This simply means that the seawater entropy (and thus $\theta_\eta$) 
is very likely the more natural conservative variable.

In addition to the examples in section 3.1 concerning the unexpected links 
between latent heat and the saturation pressure of water vapour, the examples 
described in this section show that there are observable impacts in 
geophysics of the absolute reference entropy values, thereby providing a physical 
meaning for calculations and studies of absolute seawater entropy.
However, the same doubts concerning the physical relevance of absolute 
entropy will continue to exist in atmospheric and oceanic sciences, 
much like doubts can still exist for the ultimate realism of the principles of 
many general principles of physics, such as the existence of the 
Michelson-Morley-Lorentz-Einstein's limiting velocity ($c$) of physical phenomena 
in a vacuum or the existence of the Planck's quantum of action ($h$). 
Indeed, there is no demonstration in the logical, mathematical, or physical senses 
of these facts, which are simply observed, never disproved, and therefore 
elevated to the status of general principles. 
The same applies to the third law and the principle of unattainability 
of the absolute zero of temperature, subject to amending 
\citet{Planck1911,Planck1917} 
formulation by adding possible residual entropies at $0$~K in the 
calorimetric computations for some species like \ch{H2O}, 
as logically included in $\eta_{\rm abs}$ and $\theta_\eta$ for seawater.

Accordingly, other contributions than TEOS10 have admitted the possibility 
of at least considering as an option the absolute entropies for 
liquid water and dry air \citep[ML76, M83,][]{Feistel_Hagen_1995,
Lemmon_al_2000,Feistel_Wagner_2006}. 
Nonetheless, it is often still considered that the TEOS10's choice 
is consistent with the first conclusion of the 5th International 
Conference on the Properties of Steam in London 
\citep[][]{Proc_5th_International_Conf_Prop_Steam_1956}. 
Indeed, this conference recommended (p.~1/30-1/32 and 3/34) 
that the specific internal energy and the specific 
entropy of the liquid water should be set equal to zero
at the triple point temperature of $0.01{}^{\circ}$C instead of $0.00{}^{\circ}$C, 
without affecting any measurable thermodynamic 
properties of the climate system. 
However, I show in Section~12.5 and Fig.~32 of the SM that 
this first 1956 recommendation (p.~3/34) was only valid for 
the case of a pure liquid-water steam system. 
It was not valid for a mixture involving other species, 
like in the moist-air atmosphere and the seawater. 
Moreover, it was decided at the same time 
\citep[][p.~3/35-3/37]{Proc_5th_International_Conf_Prop_Steam_1956} 
that the absolute values for the entropy of liquid water (computed from 
hypotheses made at $0$~K) should also be mentioned in all subsequent studies. 
This is precisely what I have achieved in the present study by defining 
both $\eta_{\rm abs}$ and $\theta_\eta$ for seawater, including the impacts 
of absolute and non-arbitrary definitions for both pure liquid-water 
and sea-salt reference entropies.

 \section{Conclusion}
\label{section_conclusion}
\vspace*{-2mm}

I have shown that several problems unfortunately prevent the direct use 
of the `relative' entropy formulation considered by Millero in 1976 and 1983. 
In particular, this Millero's `relative' entropy formulation did not really 
take into account the absolute values of entropies and has never been considered 
since then, and in particular in IAPWS and TEOS10 formulations.
For these reasons, I have based the present approach on the more modern 
formulation of TEOS10, with the sole addition of the absolute seawater 
entropy increment $\Delta \eta_s$ given by (\ref{Eq_etas_minus_etaw}) 
proportional to both $\eta_{\rm w0}-\eta_{\rm s0}$ and $S_{\rm A}-S_{\rm SO}$, 
with the numerical value given by (\ref{Eq_etas_minus_etaw_value}). 

The existing standard TEOS10's standard entropy formulation  
$\eta_{\rm std/TEOS10} = \eta^{\rm W}_{\rm Fei03}+\eta^{\rm S}_{\rm Fei08}$
is recalled in Tables~\ref{Table_TEOS10_sigma_W_x_y_z} and 
\ref{Table_TEOS10_sigma_S_x_y_z}. 
The reference entropies at $25{}^{\circ}$C  for 
pure liquid water ($\eta_{\rm w0}$) and sea salts ($\eta_{\rm s0}$) 
previously considered by Millero have been updated 
and computed at $0{}^{\circ}$C from more recent and 
complete thermodynamic datasets to give the values 
(\ref{eq_eta_w0_NEA_TDB_1992}) and (\ref{Eq_eta_s_0C_Marquet_2023}). 
This is based on the third law expressed by \citet{Planck1911,Planck1917},  
who generalized the heat theorem of \citet{Nernst_1906}.
Note that the residual entropy computed by \citet{Pauling1935} 
and \citet{Nagle1966} is automatically taken into account in 
the translational part of the entropy for 
liquid water, as it should be and as recalled by 
\citet[][]{Feistel_Wagner_2005,Feistel_Wagner_2006}. 

Therefore, the present paper offers the possibility of easily amending, 
if needed, the current standard formulation of TEOS10 to calculate 
the seawater absolute entropy corresponding to the recommendations 
of the third law of thermodynamics.

As a result, the entropy diagrams in 
Figs.~\ref{Fig_Sanility_parts_Seawater_Entropies} 
show large differences in magnitude (and even signs) 
for the changes in seawater entropy as far as the 
salinity is not a constant.
It should be worth adding the increment term $\Delta \eta_{\rm s}$ 
wherever salinity values are particularly low or high and wherever 
salinity gradients are large.
I show in the second part of the paper several examples 
from observations and analysed datasets where new isentropic features 
are revealed only with the absolute entropy of seawater. 
This might convince the oceanographic community, which believes 
that these differences have no physical significance, of the 
interest of the third law of thermodynamics.

Since the correction term $\Delta \eta_{\rm s}$ depends on salinity 
and thus varies not only spatially but temporally, the calculation 
of entropy itself is changed beyond the addition of a simple constant 
(like $\eta_{\rm w0}$ in $\eta^{\rm W}_{\rm Fei03}$ recalled 
in Table~\ref{Table_TEOS10_sigma_W_x_y_z}). 
Consequently, because studies of the type ``\,maximum entropy state\,'' 
defined by $d\eta = 0$ depend on a certain combination of the differentials 
$dT$ and $(\eta_{\rm w0}-\eta_{\rm s0})\:dS_{\rm A}$, there is an impact of the 
absolute reference entropy terms for non-stationary 
states with both $dT \neq 0$ and $dS_{\rm A} \neq 0$.
This may generate impacts, for example, on regional, seasonal, 
and climate changes.

Furthermore, insofar as the seawater entropy gradient regions influence 
the turbulent transport of this thermodynamic state variable, with turbulent 
flows that must cancel out in isentropic regions, the absolute definition of 
seawater entropy should impact the second principle of thermodynamics. 

Increased accuracy could be achieved by readjusting 
the tuning of the TEOS10's formulation by using the 
absolute values of both $\eta_{\rm w0}$ and $\eta_{\rm s0}$ at $0{}^{\circ}$C, rather 
than using the values $g_{010}$ and $g_{210}$ as free adjustment 
variables in Tables~\ref{Table_TEOS10_sigma_W_x_y_z} 
and \ref{Table_TEOS10_sigma_S_x_y_z}.
Nonetheless, I think that adding the missing first-order correction term 
$\Delta \eta_{\rm s}$, as done by Millero and the present paper as well, 
does not introduce major uncertainties. 
This is due to the fact that the majority of the nonlinearities have 
already been considered in the well-founded standard TEOS10 version.

Since \citet[][p.~8]{Fofonoff_1962}, 
the prevailing view in the oceanographic community is that the choice 
of the linear salinity function $a_2 +a_4\:S$ entering into the 
TEOS10 definition of the seawater entropy has no practical impact on 
known oceanographic applications and that the choice of fixing 
$a_2$ and $a_4$ is a matter of convention. 
However, some people think that the Millero and TEOS10 approaches may be  
considered valid and that neither should be subject to general rejection.
It is within this last absolute-entropy framework 
that the present and Millero's papers are situated, 
in line with the previous books of 
\citet{Fowler_Stat_Mechanics_1929},
\citet{Guggenheim_Methods_Gibbs_1933}, and 
\citet{Guggenheim_Thermodynamics_1950},  
where the third law of thermodynamics and the Sackur-Tetrode 
translational entropy were fully considered. 
Note, however, that Guggenheim rather called the absolute value of entropy 
the ``{\it conventional\:}'' entropy.   

\citet{Feistel_2019} has recently decided to take the debate 
to the related subject of the physical meaning of the residual entropy 
of ice at $0$~K.
This is a subject that is different from the translational, 
quantum degrees of freedom and the absolute (or not) status of entropy.  
This subject of residual entropy only concerns a few species such as
\ch{H2O}, \ch{N2O}, and \ch{CO}, for which proton disorder still exists  
at $0$~K.
In any case, it is clear that only by taking into account the residual 
entropy for \ch{H2O} computed by \citet{Pauling1935} and \citet{Nagle1966} 
can the calorimetric and quantum calculations of absolute entropy 
coincide, as shown in the Fig.~B1 of \citet{Marquet_Stevens_JAS22}
and as taken into account in all thermodynamic tables 
\citep[for instance, in][]{Lewis_Randall_1961,Chase_1998,
Atkins_Paula_2014,Schmidt_Thermodynamics_Engineers_Springer_2022,
Atkins_Paula_Keeler_2023}.  
Moreover, in section~3.3 I provide evidence that this residual entropy 
impacts the relationships (\ref{Eq_H20_Ls_T0}) and (\ref{Eq_H20_Psat_T0}), 
with the values of latent heat and saturated pressure becoming 
unrealistic if this residual value is not taken into account.

For my part, I have complete confidence in these third-law 
thermodynamic tables that provide the absolute entropies. 
For the more general case of a mixture of variable composition 
like seawater, and as shown in this study, there is no reason to 
continue to apply only the first pure liquid water arbitrary 
recommendations expressed (p.~3/34) at London during the   
\citet[][]{Proc_5th_International_Conf_Prop_Steam_1956}. 
It is needed to consider instead the next recommendation (p.~3/35-37), 
about the need to compute and study the absolute version of the entropy. 
Moreover, the residual entropies are small quantities and should not 
overshadow the much more fundamental aspect of the absolute values
of the entropies due to the impact of the 
translational degrees of freedom of atoms and molecules, 
as calculated by Sackur, Tetrode, and Planck in the years 
1911 to 1917.


It is true that most observable thermodynamic quantities do not 
depend on reference values of pure-water and sea-salt entropy 
(specific volumes, heat capacities, expansion coefficients, 
sound speeds, osmotic pressure, etc). 
But these reference values impact at least the seawater entropy 
itself, which deserves to be calculated and studied as one of 
the major thermodynamic state functions. 
This is confirmed by the interesting features revealed by 
the present study and the next part II.


 \section*{Acknowledgments}
\vspace*{-3mm}

I would like to thank the late Prof. Geleyn for his initial interest and 
discussions about the absolute definitions of entropy in the geophysical sciences. 

I would also like to extend my warmest thanks to Jean-Claude Andr\'e, 
whose invaluable and crucial assistance made it possible to publish 
both parts of this paper.

The TEOS10-GSW Oceanographic FORTRAN software has been downloaded from 
\url{https://www.teos-10.org/software.htm} 
\citep{McDougall_TEOS10_GSW_Getting_Started_2011}.


The author would like to thank the anonymous referees and the editor
for the constructive comments, which help to improve the manuscript.

First-Review Answers to the Editors and Reviewers 
can be found on Zenodo \citep{Marquet_Zenodo_2025_Answers_R1}.

Supplementary materials are provided in the Zenodo file
\citet{Marquet_Zenodo_2025_Sup_Mat_3rd_law}. 

Second-Review Answers to the Reviewers 
can be found on Zenodo \citep{Marquet_Zenodo_2026_Answers_R2}.

\bibliographystyle{ametsoc2014}
\bibliography{Marquet_seawater_absolute_entropy_Part1_arXiv_R2}


 \end{document}